\documentclass[11pt,a4paper]{article}
\normalsize
\usepackage[margin=2cm]{geometry}
\usepackage{graphicx}
\usepackage{amssymb,amsmath}
\usepackage{epstopdf}
\usepackage{amsthm}
\usepackage{color}
\usepackage{amsfonts}
\usepackage{amsfonts}
\usepackage{cite}
\usepackage{graphicx}
\usepackage{algorithmic}
\usepackage[center]{caption}
\usepackage{epsfig}
\usepackage{latexsym}
\usepackage{epstopdf}
\usepackage{verbatim}
\usepackage{amsthm}
\usepackage{amssymb}
\usepackage[linesnumbered,ruled,vlined]{algorithm2e}

\graphicspath{{./figures/}}
\usepackage{epstopdf}
\usepackage{amsmath}

\begin{document}
\newgeometry{left=2cm, right=2cm, top=3cm, bottom=3cm}
\title{On the Degrees of Freedom of SISO X-networks with Synergistic Alternating Channel State Information at transmitters}


\author{\large Ahmed Wagdy$^{\star\dagger}$, Amr El-Keyi$^\dagger$, Tamer Khattab$^\star$, Mohammed Nafie$^\dagger$ \\ [.1in]
\small
\begin{tabular}{c} $^\star$Electrical Engineering, Qatar University, Doha, Qatar.\\
 $^\dagger$Wireless Intelligent
Networks Center (WINC), Nile University, Egypt.\\
\end{tabular}
}

\maketitle

\begin{abstract}
In this paper, we consider the two-user single-input single-output (SISO) X-channel and $K$-user SISO X-network in fast fading environment. It is assumed that the transmitters have access to synergistic alternating channel state information (CSI). Specifically, the CSIT alternates between three states, namely, perfect, delayed and no-CSIT, in a certain manner to enable these states to work together cooperatively. These states are associated with fractions of time denoted by $\lambda_P$, $\lambda_D$  \text{and} $\lambda_N$, respectively.
For the two-user $X$-channel, simple upper bound is developed to prove the tightness of the achievability result of $4/3$ DoF under
a certain distribution of the availability of three CSIT states for $\Lambda(\lambda_P=1/3, \lambda_D= 1/3, \lambda_N=1/3)$. 
For the $K$-user $X$-network, it is shown that the sum Degrees of freedom (DoF) is at least $2K/(K + 1)$, using two-phase transmission schemes over finite symbols channel extension and under
the same distribution of the availability of  $\Lambda(\lambda_P=1/3, \lambda_D= 1/3, \lambda_N=1/3)$.This achievability result, can be considered as a tight lower bound, coincides with the best lower bound known for the same network but with partial output feedback in stead of alternating CSIT. Hence, we show that the role of synergistic alternating CSIT with distribution $\Lambda(1/3,1/3,1/3)$ is equivalent to the partial output feedback. Also, this lower bound is strictly better than the best lower bound known for the case of delayed CSI assumption for all values of $K$. 
All the proposed transmission schemes are based on two phases transmission strategy, namely, interference creation and interference resurrection, which exploit the synergy of instantaneous CSI and delay CSIT to retrospectively align interference in the subsequent channel uses. The proposed transmission schemes offer DoF gain compared to partial output feedback, delayed CSIT and no-CSIT. The achievable DoF results are the best known results for these channels.


\end{abstract}


\section{Introduction}
The scarcity of the wireless spectrum and the increasing growth of high data rate demands arise the impossibility of separating
the concurrent transmission completely in frequency and impose that the transmissions necessarily occur at the same time in the same frequency band, separated only in space which introduce more signal interference in wireless networks. As a result, the received signal at   each receiver is the desired transmitter’s signal of intended user in addition to the signals from many undesired or interfering transmitters of the other users. Consequently, it is widely known that signals interference is the main performance limiting factor of most wireless networks. Moreover, as the number of users in a wireless network sharing the same spectrum increases, the network becomes more and more interference limited. Therefore, establishing the performance limits of wireless networks turns out to be more challenging. 

Interference alignment, the state-of-the-art in interference management\cite{jafar2011interference}, arises the possibility of establishing the performance limits of wireless networks in terms of characterizing the sum degrees of freedom  (DoF) of many wireless networks. For example, in \cite{X-Networks} it was shown that M$\times$N X-network can achieve $\frac{MN}{M+N-1}$ DoF, the DoF upper bound of that network, using simple interference alignment scheme over infinite symbols channel extension. The K-user SISO X-channel is the most comprehensive and fundamental setting for the information theoretic study of interference alignment in multi-user wireless networks. Interestingly, this setting, can be considered as a
combination of broadcast and multi-access channels, is the general case where each transmitter has an independent message for each receiver, for a total of $K^2$ independent messages. Specifically, if all the transmitters are combined into one compound transmitter with $K$ transmit antennas and the all the receivers are combined into one compound receiver with $K$ receive antennas, then the resulting setting is point to point MIMO channel. While, if transmitters are combined into one compound transmitter and the receivers are remained distributed, then the resulting setting is MISO Broadcast channel. Also, the interference channel and Z-channel can be derived out from the X-channel by setting the appropriate messages to null. Moreover, the studying of the X-channel was contributed to showing the great of potential of the interference alignment in the early stages. In particular, the authors of \cite{Signaling}introduce an efficient signaling scheme, known as MMK scheme, which works at a corner point of the achievable region for the MIMO X-channel, specifically, it is shown that for a MIMO X channel with $3$ antennas at each node, a degrees of freedom of $4$ is
achievable by a combination of dirty paper coding, successive decoding and zero forcing. While in \cite{MIMO-X} it is proved that simple zero forcing is sufficient to achieve the same achievability results of the MMK scheme in \cite{Signaling}, the key idea for this interesting approach is interference alignment. 

Considerable work in literature of interference alignment has focused on characterizing the degrees of freedom of X-channel and X-network. Contrary to what has been established in the context of memory-less point-to-point channel that the channel feedback does not increase the capacity \cite{Shannon}, the channel feedback, known as CSIT, in multi-user networks can significantly widen the capacity region and hence the degrees of freedom region. Throughout literature, the CSIT plays a leading rule in charactrizing DoF of wireless networks, and was the canonical motif and the influential ingredient in developing the phenomenal interference alignment techniques. Under full CSIT assumption; where the transmitters have global and instantaneously perfect CSI, the wireless networks achieve the highest DoF and enjoy the widest DoF region. In \cite{X-Networks}, it is proved that the DoF of M$\times$N-user SISO X-network with full CSIT is upper bounded by $\frac{MN}{M+N-1}$ also the authors proposed a partial interference alignment scheme that asymptotically approach the upper on DoF within an $\epsilon > 0$ by considering large channel extensions. In certain cases, when whether the number of transmitters or receivers are equal to two, the upper bound is achievable and perfect interference alignment is attained within finite channel extension. On the other hand, in the total lack of CSIT, the DoF region of most wireless networks collapse to narrowest region, where it's corner points are achievable simply  by time or frequency division multiplexing between users \cite{huang2012degrees, No-CSIT} however, in certain scenarios, the interference alignment is still feasible. Specifically, in \cite{blind}, Syed Jafar paved the way to achieve interference alignment by exploiting only the knowledge of heterogeneous channel coherence structures associated with different users in the same network even in complete lack of knowledge of the channel at the transmitters, i.e. the X channel without no CSIT and under the heterogeneous block fading in both time and frequency assumption; one user suffers time selectivity and other is frequency-selective, achieves $4/3$ DoF and hence coincide with the best known DoF upper bound on it. 

Extensive research efforts have been devoted to discover middle ground between the two extremes; full CSIT and no CSIT, such as quantized CSIT \cite{Finite_rate, Finite_rate_Kobayashi}, compound CSIT \cite{weingarten2007compound, gou2011degrees, maddah2010degrees} and others that make use of temporal correlation yet the most remarkable one is what is widely known as delayed CSIT. This model was first introduced by Maddah Ali and David Tse in \cite{completely} for the Gaussian multiple-input single-output(MISO)broadcast channel
(BC). The delayed model introduced  a fundamental, and rather counter-intuitive observation that the completely outdated channel knowledge to the transmitters in i.i.d. Rayleigh fading model,where the channels take completely independent values every time slot, creates great opportunities for interference alignment and significantly improve the DoF of MISO BC. In \cite{maleki2012retrospective}, Maleki et al.  applied the delayed CSIT model to the distributed transmitters networks such as X channel and interference channel. They showed that 
the 2-user SISO X channel and 3-user SISO interference channel, under delayed CSIT assumption, can achieve $8/7$ and $9/8$ DoF, respectively. Then, in \cite{ghasemi2011degrees} and et al. introduced  a new transmission strategy specially tailored to the distributed transmitters networks that efficiently exploited the delayed channel knowledge to provide new achievability results that outperforms what have been obtained in \cite{maleki2012retrospective}. Particularly, they showed that $6/5$ and $5/4$ is achievable for the 2-user and 3-user X-channel, respectively. 
In this paper, We consider a two user Gaussian X channel where
each node is equipped with single antenna. In this channel,
transmitters $T_1$ and $T_2$ have four independent messages
$W_{11}$, $W_{12}$, $W_{21}$ and $W_{22}$ for receivers $R_{1}$
and $R_{2}$ such that $W_{ij}$ originates at transmitter $j$ and
is intended for receiver $i$. Earlier research work on the DoF of
the two-user X channel have determined that the upper bound for
DoF of two-user single-input single-output (SISO) X channel is
$4/3$ and for MIMO one is $4M/3$ where $M$ is the number of
antennas per node \cite{maddah2008communication, 5208535,4418479}.
These upper bounds are achievable with global, perfect and
instantaneous CSIT and when the channel coefficients are time
varying or frequency selective and drawn from continuous
distribution. The authors of \cite{maleki2012retrospective} showed
that even in fast fading environment and for interference networks
consisting of distributed transmitters and receivers, delayed CSIT
channel could be beneficial and have a great impact on increasing
DoF. They proved that for the two-user SISO X channel, the $8/7$
DoF is achievable with delayed CSIT. New results have been
demonstrated in \cite{ghasemi2011degrees} where the two user SISO
X channel with delayed CSIT could achieve $6/5$ DoF and for three
user X network could achieve $5/4$ DoF.\\

In this work, we consider the two-user SISO X channel with
alternating CSIT. The main question we ask is whether the
\textit{synergistic alternation} in the availability of CSIT  is
beneficial in this channel as it is in collocated transmitters
networks. We answer this question in an affirmative way by
presenting Theorem 1 and discussing the synergistic benefits.
Unlike what is commonly thought that the synergistic benefits of
alternating CSIT could be more sensitive to or may be lost
depending on whether the transmitter of the network are
distributed or collocated, we end up with the synergistic
alternation of CSIT is still very beneficial in distributed
transmitters network. The second question we ask is any alteration
pattern is synergistic alternation and has synergistic benefits
that could provide extraordinary DoF gain. We negatively answer
this question through remarks 1,2,3 and 4; where, we show that
there exist some certain alteration patterns in which the
different channel knowledge availability states could not work
together in cooperative way but they work individually and the
corresponding DoF dwindle to the sum of their individual DoF.
Furthermore we find the synergistic alternation patterns and
dissociative ones.

\section{System Model}
A $K$-user SISO X channel is considered. In this channel, there
are $K$ independent transmitters $T_1$,... $T_K$ communicating
with $K$ independent receivers $R_1$,... $R_K$, where each node is
equipped with a single antenna. Each transmitter has an
independent message for each receiver. The received signal at the
$i$th receiver at time slot $t$ is given by
\begin{equation}\label{eq1}
Y_i(t)= \displaystyle\sum_{j=1}^K h_{ij}(t)X_j(t)+N_i(t),
\end{equation}
where $X_j(t)$ is the transmitted signal from $T_j$ at the $t$th
time slot which satisfies the power constraint
$E\{|X_{j}(t)|^2\}\leq P_{j}$. The noise $N_i(t) \sim
\mathcal{CN}(0,1)$ is the circularly symmetric complex additive
white Gaussian noise with zero mean and unit variance generated at
$R_i$ at time slot $t$. In \eqref{eq1}, $h_{ij}(t)$ is the channel
coefficient from $T_j$ to $R_i$ and all channel coefficients are
independent identically distributed (i.i.d.) over time and drawn
from a continuous distribution.

\begin{figure}
  \centering
\includegraphics[scale= 1.2]{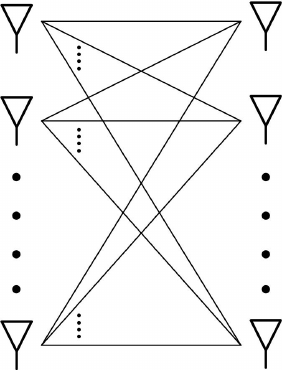} 
\caption{ $K$-user SISO X network } 
\end{figure}

We assume that the receivers know all the channel coefficients
instantaneously and with infinite precision thus global and
perfect CSI is assumed at the receivers. In contrast, we consider
three different states of the availability of CSIT; perfect (P),
delayed (D), and no-CSIT (N). These states denote respectively the
availability of CSIT instantaneously and without error, with some
delay $\geq$ one time slot and without error, and the
unavailability of CSIT at all. Let the state of CSIT availability
of the channels to the $i$th receiver be denoted by $S_i$; where,
$S_i$ $\in \{P,D,N\}$, i.e., $S_2= P$ indicates that each transmitter $j$, where $j\in\lbrace 1,2,\ldots, K\rbrace$, has perfect and instantaneous knowledge of $h_{2j}$. In addition, let
$S_{1 \ldots K}$ denote the state of CSIT availability for the channels to the network;the first, second, $\ldots$ $K^{th}$ receiver, respectively. Therefore, $S_{1 \ldots K} \in \{PP\cdots P, PP\cdots D,\ldots, NN\cdots N\}$. For example, $S_{123}= PDN$
refers to the case where $T_j$ has perfect knowledge of $h_{1j}$, delayed knowledge of $h_{2j}$, and no information about $h_{3j}$.
Moreover, we denote the CSIT availability of the channels to
receiver $i$ over $n$ time slots of time channel extension by
$n$-tuple $S_i^n =(S_i(1),\ldots, S_i(n))$. Similarly, the
availability of CSIT for the channels to the network over $n$ time slots channel extension, known by
\textquotedblleft CSIT pattern\textquotedblright, is denoted by
$S_{1 \ldots K}^n=(S_{1 \cdots K}(1),\ldots, S_{1 \ldots K}(n))$.

The fraction of time associated with the state of CSIT
availability for the network, denoted by $\lambda_{S}$ where $S
\in \{ P, D, N\}$ is given by
\begin{equation}
\lambda_{S}=\frac{\displaystyle\sum_{t=1}^n\displaystyle\sum_{i=1}^k
\mathbb{I}_S(S_i(t))}{nk},
\end{equation}
where $\mathbb{I}$ denotes the Indicator function and $k$ is the
number of users, and hence,
\begin{equation}
\sum_{S=P,D,N}\lambda_{S}=1.
\end{equation}
Furthermore, we use $\Lambda(\lambda_P, \lambda_D, \lambda_N )$ to
denote the distribution of the fraction of time for the different
states $\{P, D, N\}$ of CSIT availability.

Let $r_{ij}(P)= \frac{\log_2(|W_{ij}|)}{n}$ denote the rate of
$W_{ij}$ for a given transmission power $P$ where $|W_{ij}|$
denotes the size of the message set and $n$ is the number of
channel uses. The rate $r_{ij}(P)$ is achievable if there exists a
coding scheme such that the probability of error in decoding
$W_{ij}$ goes to zero as $n$ goes to infinity for all $(i,j)$. The
DoF region $\mathcal{D}(\Lambda)$ is defined as the set of all
achievable tuples $(d_{11},\ldots, d_{1K},d_{21},\ldots, d_{2K},\ldots, d_{K1},\ldots,d_{KK})\in \mathbb{R}_+^{K^2}$, where
$d_{ij}= \lim_{P \rightarrow \infty}
\frac{R_{ij}(P)}{\log_2{(P)}}$ is the DoF for message $W_{ij}$.
The DoF of the network is defined as:
\begin{equation}
\text{DoF}(\Lambda)= \max_{(d_{11},\ldots,d_{22}) \in  \mathcal{D}(\Lambda)} \displaystyle\sum_{j=1, i=1}^K d_{ij}.
\end{equation}

\section{Two-user SISO X Channel} \label{section_schems}
Motivated by the fact that multiuser networks with time varying channels, the variation in the availability of CSIT or
the fluctuation in state of CSIT across different links is
inevitable, we extend this verity modelled for MISO broadcast
channel in \cite{6471826} to the two-user SISO X channel. Form
\cite{maddah2008communication, 5208535,4418479}, it is known that
for the two-user SISO X channel the DoF of the network are bounded
by $4/3$. This upper bound is achievable over $3$-symbol channel
extension if the channel coefficients vary over time 
and each transmitter has global and \textit{constantly} perfect
CSIT over the $3$ time slots.

\begin{figure}
  \centering
\includegraphics[scale=.7]{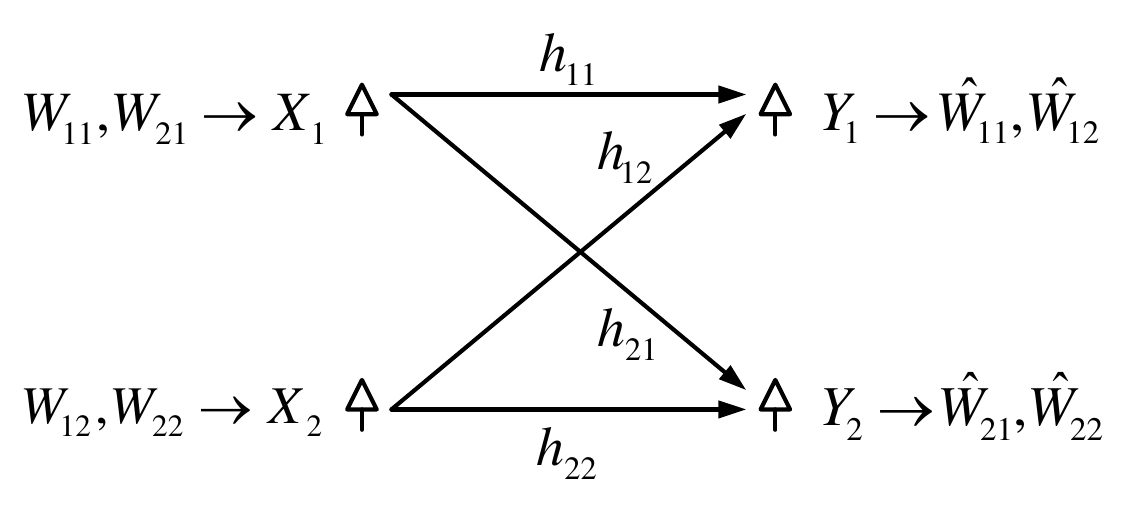} 
\caption{Two-user SISO X channel } \label{2user}
\end{figure}

In this section, we characterize the degree of freedom  region of the two-user SISO x-channel, depicted in Fig. \ref{2user}, with synergistic alternating CSIT, specifically,we present three illustrative examples for the proposed achievable scheme in three different patterns of CSIT availability as well as the converse proof. In all these cases, we show that $4/3$ DoF is achievable by sending $4$ different data symbols; $2$ for each receiver over three time slots. The basic idea behind the proposed achievable scheme is to resurrect the interference formerly created, hence, interference creation-resurrection (ICR) strikes and  interference alignment arises.

Inspired by the MAT scheme in \cite{maddah2012completely}, the proposed achievable scheme is performed in two phases over three time slots. The first phase is associated with the delayed CSIT state where
the transmitters transmit their messages. As a result, the
receivers get linear combinations of their desired messages in
addition to interference. This phase is called ``interference
creation'' phase. On the other hand, the second phase is
associated with the perfect CSIT state and is called the
``interference resurrection'' phase. In this phase, the
transmitters reconstruct the old interference by exploiting the
delayed CSIT received in phase one and the perfect CSIT in the
second phase. Hence, after three time slots, each receiver has two
different linear combinations of its desired messages and only one
interference term received twice. Noteworthy, in some cases the
two phases can overlap over the 3 time slots.

\subsection{Achievability Schemes} \label{TwoUserSchemes}
Let $u_1$ and $u_2$ be two independent symbols intended to $R_1$
transmitted from $T_1$ and $T_2$, respectively. Also, let $v_1$
and $v_2$ be two independent symbols intended to $R_2$ from $T_1$
and $T_2$, respectively. In the next subsections, we show that we
can reliably transmit these  symbols to their target destinations
in 3 time slots in three different cases of alternating CSIT.

\subsubsection{Scheme 1: Combined delayed and distributed perfect CSIT}
As an illustrative example of this case, let us consider a 2-user
SISO X channel with alternating CSIT  pattern given by (DD, PN,
NP) over three time slots. Here, we have combined delayed CSIT in
the first time slot and distributed perfect CSIT over the last two
time slots. Consequently, the proposed scheme is performed in two
separate phases as follows.

\textit{Phase one:} In this phase, the two transmitters greedily
transmit all data symbols, i.e., $X_1(1)= u_1 + v_1$ and $X_2(1)=
u_2 + v_2$. As a result, the received signals are given as
\begin{eqnarray}
Y_1(1)&\!\!\!=\!\!\!& h_{11}(1)u_1 + h_{12}(1)u_2+ h_{11}(1)v_1+h_{12}(1)v_2 \nonumber \\
&\!\!\!\equiv \!\!\!& L_1^1(u_1,u_2)+ I_1(v_1,v_2)\\
Y_2(1)&\!\!\!=\!\!\!&h_{21}(1)u_1 + h_{22}(1)u_2+ h_{21}(1)v_1+h_{22}(1)v_2 \nonumber \\
&\!\!\!\equiv\!\!\!& I_2(u_1,u_2)+ L_2^1(v_1,v_2)
\end{eqnarray}
where $L_i^j(x_1,x_2)$ denotes the $j$th linear combination of the
two messages $x_1$ and $x_2$ that are intended for receiver $R_i$
and $I_i(z_1,z_2)$ denotes the interference term for
receiver $R_i$ which is a function of the messages $z_1$ and $z_2$
that are not intended for this receiver.

\textit{Phase two:} This phase consists of two time slots
where in each time slot the transmitted signals are designed such
that the interference is resurrected at one receiver while the
second receiver receives a new linear combination of its desired
messages. Note that now the transmitters are aware of the CSIT of
the previous time slot, i.e., $T_1$ knows $h_{11}(1)$ and
$h_{21}(1)$ while $T_2$ knows $h_{12}(1)$ and $h_{22}(1)$. Also,
at $t=2$, the channels to the first receiver are known perfectly
and instantaneously at the two transmitters, i.e., $T_1$ knows
$h_{11}(2)$ and $T_2$ knows $h_{12}(2)$. As a result, the first
time slot in this phase is dedicated to resurrecting the
interference $I_1(v_1,v_2)$ received by $R_1$ in the first time
slot. The transmitted signals of $T_1$ and $T_2$ are given by
\begin{eqnarray}
X_1(2)&\!\!\!=\!\!\!&h_{11}^{-1}(2)h_{11}(1)v_1\\
X_2(2)&\!\!\!=\!\!\!&h_{12}^{-1}(2)h_{12}(1)v_2
\end{eqnarray}
and the received signals at $R_1$ and $R_2$ are given respectively
by
\begin{eqnarray}
Y_1(2)&\!\!\!=\!\!\!& h_{11}(1)v_1+h_{12}(1)v_2 \equiv
I_1(v_1,v_2)\\
Y_2(2)&\!\!\!=\!\!\!&h_{21}(2)h_{11}^{-1}(2)h_{11}(1)v_1+h_{22}(2)h_{12}^{-1}(2)h_{12}(1)v_2 \nonumber \\
&\!\!\!\equiv \!\!\!&L_2^2(v_1,v_2)
\end{eqnarray}
Hence, at the end of sub-phase one, $R_1$ has received
$I_1(v_1,v_2)$, the interference received in the first time
slot, and $R_2$ has received a new linear combination
$L_2^2(v_1,v_2)$.

In the second sub-phase, the transmitted signals are designed to
resurrect the interference received by $R_2$ in the first time
slot and provide a new linear combination of the desired messages
to $R_1$. The transmitted signals of $T_1$ and $T_2$ are given
respectively by
\begin{eqnarray}
X_1(3)&\!\!\!=\!\!\!&h_{21}^{-1}(3)h_{21}(1)u_1\\
X_2(3)&\!\!\!=\!\!\!&h_{22}^{-1}(3)h_{22}(1)u_2
\end{eqnarray}
where $T_1$ and $T_2$ utilize their perfect and instantaneous
knowledge of their channels to $R_2$.  The received signals at
$R_1$ and $R_2$ are given by:
\begin{eqnarray}
Y_1(3)&\!\!\!=\!\!\!&h_{11}(3)h_{21}^{-1}(3)h_{21}(1)u_1
+h_{12}(3)h_{22}^{-1}(3)h_{22}(1)u_2 \nonumber\\&\!\!\!\equiv
\!\!\!&
L_1^2(u_1,u_2) \\
Y_2(3)&\!\!\!=\!\!\!& h_{21}(1)u_1+h_{22}(1)u_2 \equiv
I_2(u_1,u_2)
\end{eqnarray}

After the third time slot, the two receivers $R_1$ and $R_2$ have
enough information to decode their intended messages. In
particular, $R_1$ has access to two different equations in $u_1$
and $u_2$ only. The first one is obtained by subtracting $Y_1(2)$
from $Y_1(1)$ to cancel out the interference and the second
equation is $Y_1(3)$ by itself as it is received without
interference. Similarly, $R_2$ forms its first equation by
subtracting $Y_2(3)$ from $Y_2(1)$ to cancel out the interference
while the second equation is $Y_2(2)$.

Note that this scheme could be used also when the CSIT  pattern
given by (DD, NP, PN) but with minor modification in phase two,
where sub-phase one is dedicated to resurrecting interference of
$R_2$ instead of resurrecting interference of $R_1$ and sub-phase
two is dedicated to resurrecting interference of $R_1$ instead of
resurrecting the interference of $R_2$.

\subsubsection{Scheme 2: Distributed delayed and combined perfect CSIT}
Let us consider the 2-user SISO X channel with alternating CSIT
given by (ND, DN, PP). Unlike case 1, here we have distributed
delayed CSIT over the first two time slots and combined perfect
CSIT in the last time slot. Consequently, the  interference
creation phase extends over two time slots while the interference
resurrection phase can be executed in one time slot as follows.

\textit{Phase one:} Each time slot of this phase is dedicated to
one receiver where the two transmitters transmit the desired
messages for this receiver. For example, if the first time slot is
dedicated to $R_1$, then $T_1$ transmits $u_1$ and $T_2$ transmits
$u_2$. The received signals at $R_1$ and $R_2$ are given
respectively by
\begin{eqnarray}
Y_1(1)&\!\!\!=\!\!\!&h_{11}(1)u_1 + h_{12}(1)u_2 \equiv
L_1^1(u_1,u_2)\\
Y_2(1)&\!\!\!=\!\!\!&h_{21}(1)u_1 + h_{22}(1)u_2 \equiv
I_2(u_1,u_2)
\end{eqnarray}
Therefore, $R_1$ receives linear combination $L_1^1(u_1,u_2)$ of
its desired signals, while $R_2$ receives only interference
$I_2(u_1,u_2)$. Similarly, in the next time slot, $T_1$
transmits $v_1$ and $T_2$ transmits $v_2$ and the received signals
at $R_1$ and $R_2$ are given respectively by
\begin{eqnarray}
Y_1(2)&\!\!\!=\!\!\!&h_{11}(2)v_1 + h_{12}(2)v_2 \equiv I_1(v_1,v_2)\\
Y_2(2)&\!\!\!=\!\!\!&h_{21}(2)v_1 + h_{22}(2)v_2 \equiv
L_2^1(v_1,v_2)
\end{eqnarray}
where  $R_2$ receives linear combination $L_2^1(v_1,v_2)$ of its desired signals, while $R_1$ receives only interference $I_1(v_1,v_2)$.

\textit{Phase two}: This phase includes only one time slot where the
transmitters resurrect the formerly received interference terms
$I_1(v_1,v_2)$ and $I_2(u_1,u_2)$, while providing new linear
combinations of the desired messages to the two receivers. In
order to achieve this goal, the transmitted signals from $T_1$ and
$T_2$ in the third time slot is given by
\begin{eqnarray}
X_1(3)&\!\!\!=\!\!\!&h_{21}^{-1}(3)h_{21}(1)u_1+h_{11}^{-1}(3)h_{11}(2)v_1\\
X_2(3)&\!\!\!=\!\!\!&h_{22}^{-1}(3)h_{22}(1)u_2+h_{12}^{-1}(3)h_{12}(2)v_2
\end{eqnarray}
and the corresponding received signals at $R_1$ and $R_2$ are
given respectively by
\begin{eqnarray}
Y_1(3)&\!\!\!=\!\!\!& L_1^2(u_1,u_2)+ I_1(v_1,v_2)\\
Y_2(3)&\!\!\!=\!\!\!& L_2^2(v_1,v_2)+ I_2(u_1,u_2)
\end{eqnarray}
where {\small
\begin{eqnarray}
 L_1^2(u_1,u_2)&\!\!\!\!=\!\!\!\!& h_{11}(3)h_{21}^{-1}(3)h_{21}(1)u_1
\!+\!h_{12}(3)h_{22}^{-1}(3)h_{22}(1)u_2 \qquad\\
L_2^2(v_1,v_2)&\!\!\!\!=\!\!\!\!&
h_{21}(3)h_{11}^{-1}(3)h_{11}(2)v_1\!+\!h_{22}(3)h_{12}^{-1}(3)h_{12}(2)v_2\qquad
\end{eqnarray} } \normalsize
At the end of the third time slot, each receiver can decode its
intended messages by solving two equations. For example, $R_1$
subtracts $Y_1(2)$ from $Y_1(3)$ to cancel out the interference
and obtain the first equation in $u_1$ and $u_2$ while the second
equation is $Y_1(1)$ by itself as it received without
interference.

Noteworthy, this scheme could be used when the CSIT pattern is
given by (DN, ND, PP) but with minor modification in phase one
where the two sub-phases swap their dedications from $R_1$ to
$R_2$ and vise versa.

\subsubsection{Scheme 3: Distributed delayed and distributed perfect CSIT}
As an illustrative example, let us consider a 2-user SISO X
channel with CSIT pattern given by (DN, PD, NP). Unlike the above
two examples, we have distributed delayed CSIT over the first two
time slots and distributed perfect CSIT over the last two
consecutive time slots. Consequently, the proposed scheme is
performed in two overlapping phases as follows.

\textit{Time slot 1:} The first sub phase of phase one begins at
$t=1$, and is dedicated to transmitting the desired messages of
$R_2$, i.e., $T_1$ transmits $v_1$ while $T_2$ transmits $v_2$.
The received signals are this given by
\begin{eqnarray}
Y_1(1)&\!\!\!=\!\!\!&h_{11}(1)v_1 + h_{12}(1)v_2 \equiv
I_1(v_1,v_2) \\
Y_2(1)&\!\!\!=\!\!\!&h_{21}(1)v_1 + h_{22}(1)v_2 \equiv
L_2^1(v_1,v_2)
\end{eqnarray}
Therefore, $R_2$ receives the first linear combination $L_2^1(v_1,v_2)$ of its desired signals, while $R_1$ receives only interference $I_1(v_1,v_2)$.

\textit{Time slot 2}: At $t=2$ the overlap occurs between the two
phases. In particular, sub-phase two of phase one and sub-phase
one of phase two begin simultaneously. In this time slot,
sub-phase two of phase one creates interference at $R_2$ with
while sub-phase one of phase two is designed to resurrect the
interference term $I_1(v_1,v_2)$. The transmitted signals are
given by:
\begin{eqnarray}
X_1(2)&\!\!\!=\!\!\!&u_1+h_{11}^{-1}(2)h_{11}(1)v_1 \\
X_2(2)&\!\!\!=\!\!\!&u_2+h_{12}^{-1}(2)h_{12}(1)v_2
\end{eqnarray}
and the corresponding received signals are given by:
\begin{eqnarray}
Y_1(2)&\!\!\!=\!\!\!&h_{11}(2)u_1 + h_{12}(2)u_2 +h_{11}(1)v_1 + h_{12}(1)v_2 \nonumber \\
&\!\!\!\equiv\!\!\!&  L_1^1(u_1,u_2)+I_1(v_1,v_2)\\
Y_2(2)&\!\!\!=\!\!\!& h_{21}(2)h_{11}^{-1}(2)h_{11}(1)v_1+h_{22}(2)h_{12}^{-1}(2)h_{12}(1)v_2 \nonumber \\
&\!\!\!+\!\!\!&h_{21}(2)u_1 + h_{22}(2)u_2\nonumber \\
&\!\!\!\equiv\!\!\!& L_2^2(v_1,v_2)+I_2(u_1,u_2)
\end{eqnarray}
Therefore, $R_2$ receives a new linear combination
$L_2^2(v_1,v_2)$ of its desired signals and an interference term
$I_2(u_1,u_2)$ as a by-product of the overlap, while $R_1$
receives the old interference $I_1(v_1,v_2)$ and the first linear
combination $L_1(u_1,u_2)$ of its desired signals.

\textit{Time slot 3}:  In this time slot the  transmitters send
linear combination from $u_1$ and $u_2$ aiming to resurrect the
interference terms $I_2(u_1,u_2)$ formerly received at $t=2$,
while providing a new linear combinations to $R_1$ of its desired
messages. The transmitted signals are given by:
\begin{eqnarray}
X_1(3)&\!\!\!=\!\!\!&h_{21}^{-1}(3)h_{21}(2)u_1\\
X_2(3)&\!\!\!=\!\!\!&h_{22}^{-1}(3)h_{22}(2)u_2
\end{eqnarray}
and the corresponding received signals are
\begin{eqnarray}
Y_1(3)&\!\!\!=\!\!\!&h_{11}(3)h_{21}^{-1}(3)h_{21}(2)u_1
+h_{12}(3)h_{22}^{-1}(3)h_{22}(2)u_2\nonumber \\
&\!\!\!\equiv\!\!\!& L_1^2(u_1,u_2)\\
Y_2(3)&\!\!\!=\!\!\!&h_{21}(2)u_1+h_{22}(2)u_2 \equiv
I_2(u_1,u_2)
\end{eqnarray}
Finally, the two receivers $R_1$ and $R_2$ have enough information
to decode their intended messages.  In particular, $R_1$ has
access to two different equations in $u_1$ and $u_2$ only. The
first one is obtained by subtracting $Y_1(1)$ from $Y_1(2)$ to
cancel out the interference and the second equation is $Y_1(3)$ by
itself as it is received without interference. Similarly, $R_2$
its first equation is $Y_2(1)$ while forming its second equation
by subtracting $Y_2(3)$ from $Y_2(2)$ to cancel out the
interference.

Note that this scheme could be used when the CSIT pattern is given
by (ND, DP, PN)  but with minor modification in the two phases
where the two sub-phases in each phase swap their dedications from
$R_1$ to $R_2$ and vise versa .
\subsection{Synergistic CSIT Alternation Patterns}
In this section, we discuss CSIT alternation patterns that can
provide synergistic gain in the DoF of the two-user SISO X
channel within three-symbol channel extension.
Since the possible CSIT states for the two users are given by
$S_{12}\in \{\text{PP,PD,PN,DP,DD,DN,NP,ND,NN}\}$, there are $9^3$
possible alternation patterns over the three time slots.

First, we note that the aforementioned three examples in Section
\ref{section_schems} present the CSIT patterns with the lowest
CSIT sufficient to achieve $4/3$ DoF. Definitely, any alternation
pattern with channel knowledge higher than these patterns can
achieve the same DoF, i.e., if we have $S_{12}= \text{ND}$, its
higher CSIT state that could provide the same synergistic DoF gain
are $\{\text{NP, DD, DP, PD, PP}\}$. Theorem 1 presents sufficient
conditions on the lowest CSIT alternation pattern among three-symbol 
channel extension patterns for achieving
the upper bound on the DoF of the two-user SISO X channel.\\

\textit{Theorem 1:} For the two-user SISO X channel in time
varying or frequency selective settings, the upper bound on the
total DoF of the channel is achievable if the following 
requirements on the CSIT alternation pattern are satisfied.
\begin{enumerate}
    \item Each transmitter has a delayed CSIT followed by a perfect CSIT over three time slots.
    \item At each time slot, at least one transmitter should have some CSIT (perfect or delayed),
    i.e., the two transmitters should not be simultaneously
    without CSIT.
    \item In the third time slot, at least one transmitter should
    have perfect CSIT.\\
\end{enumerate}

\textit{Proof:} We show that the three requirements of Theorem 1
limit the CSIT alternation patterns in
 a 3-symbol channel extension to the minimum CSIT
synergistic patterns considered in the three examples of Section
\ref{section_schems}, and its higher CSIT patterns. Hence, the
achievability of $4/3$ DoF follows from the results of Section
\ref{section_schems}. The first requirement in Theorem 1 yields
three possible minimum states for the CSIT of the channel to the
$i$th receiver over three time slots $S_i^T \in \{\text{(D,P,N),
(D,N,P), (N,D,P)}\}$. As a result we have 9 possible combinations
for the CSIT of the two-user channel. Six of these 9 combinations,
satisfy the second and third requirements in Theorem 1 and are
listed as the first 6 entries in Table 1. The remaining three
combinations are those which have $S_{12}= \text{NN}$ in any of
the three time slots, i.e., (DD,PP,NN), (DD,NN,PP), and
(NN,DD,PP). For the first combination, the minimum CSIT states
that satisfy the three requirements are (DD,PP,PN), which is
higher than (DD,NP,PN), and (DD,PP,NP), which is higher than
(DD,PN,NP). From Table 1, we can achieve $4/3$ DoF using
achievable scheme 1 in both cases. Similarly, for the CSIT state
(DD,NN,PP), the minimum CSIT that satisfy the requirements of
Theorem 1 are (DD,ND,PP) and (DD,DN,PP) for which $4/3$ DoF can be
achieved using scheme 2. Finally, for the CSIT state (NN,DD,PP),
the minimum CSIT that satisfy the requirements of Theorem 1 are
(ND,DD,PP) and (DN,DD,PP) for which $4/3$ DoF can be achieved
using scheme 2
too. \\

\begin{table}[!ht]
\centering
\begin{tabular}{|l|p{2cm}||l|p{2cm}|}
  \hline
   CSIT state & Scheme  &CSIT state &  Scheme  \\
  \hline
 $(DD,PN,NP)$ & Scheme 1 &$(DN,PD,NP)$ & Scheme 3\\
 $(DD,NP,PN)$ & Scheme 1 &$(DN,ND,PP)$ & Scheme 2\\
 $(ND,DP,PN)$ & Scheme 3 &$(ND,DN,PP)$ & Scheme 2\\
  \hline
\end{tabular}\\
\caption {Achievable schemes for different CSIT states}
\end{table}

\subsection{The Degrees of Freedom Region}
In this section, we characterize the degrees of freedom region $\mathcal{D}(\Lambda)$ and the sum degrees of freedom $DoF(\Lambda)$ for the SISO X channel with alternating CSIT for a certain region of distributions illustrated in Fig.1 where $\Lambda( \lambda_P\geq 1/3, \lambda_D\geq 2/3-\lambda_P)$.

\subsubsection{Outer bound}
\textit{Theorem 2:} $\mathcal{D}(\Lambda)\subset \mathcal{D}_{out}(\Lambda)$, where $\Lambda( \lambda_P\geq 1/3, \lambda_D\geq 2/3-\lambda_P)$ and the outer bound on the degrees of freedom is defined as followed: 
\begin{eqnarray}
\mathcal{D}_{out}(\Lambda)\triangleq \bigg \{ \left( d_{11}, d_{12},d_{21},d_{22}\right)  \in \mathbb{R}_+^4:\\ \nonumber
&& d_{11}+d_{12}+d_{21}\leq 1\\ \nonumber 
&& d_{11}+d_{12}+d_{22}\leq 1\\ \nonumber
&& d_{11}+d_{21}+d_{22}\leq 1\\ \nonumber 
&& d_{12}+d_{21}+d_{22}\leq 1\bigg \}. 
\end{eqnarray}

\begin{proof}
    We consider the SISO X channel with perfect CSIT as a special case of X channel with alternating CSIT where the CSIT availability  distribution is  $\Lambda=\big(\lambda_{P}=1, \lambda_{D}=0,\lambda_{N}=0 \big)$. Definitely, That case has the highest channel knowledge available to the transmitters that they need to align interference and deliver their messages perfectly to their intended receivers. While, additional channel knowledge cannot deteriorate the upper bound, it could make it quite loose, however in our case, we show in \ref{DOF_Region} that the upper bound is tight. Therefore, the upper bound on the degrees of freedom with perfect availability of CSIT, $\Lambda=\big(\lambda_{P}=1, \lambda_{D}=0,\lambda_{N}=0 \big)$, is strictly an outer bound for the degree of freedom of SISO X channel with alternating CSIT given any certain distribution of the availability CSIT. We can consider the SISO X channel as four different interlocking SISO Z channels; each one is formed by eliminating one message and setting the corresponding channel between any transmitter-receiver pair to zero i.e. $Z_{12}$ is the conventional X channel but with $h_{12}=0$. Calling the results in \cite{5208535}, in the context of Z channel,specifically, Lamma 1, stating that the maximum sum of degrees of freedom of a SISO Z channel over the degrees of freedom region of Z channel is an upper bound to the maximum sum of the degrees of freedom the X channel over the corresponding degrees of freedom region of X channel. In addition, Corollary 1, stating that the maximum sum of degrees of freedom of the X channel over the degrees of freedom region of corresponding  SISO Z channel is upper bounded by one. Therefore, the maximum sum of any three degrees of freedom, corresponding to different Z channel, over the degrees of freedom region of SISO X channel is upper by one. Hence, The four conditions corresponding to the four different Z channels, represent outer bounds of the SISO X channel with perfect CSIT, and are straightforward outer bounds for the SISO X channel with alternating CSIT under any distribution $\in\Lambda( \lambda_P\geq 1/3, \lambda_D\geq 2/3-\lambda_P)$. 
\end{proof}

\subsubsection{The Degrees of Freedom Region} \label{DOF_Region}
\textit{Theorem 3:}
The degrees of freedom region of the two user SISO X channel with alternating CSIT under any distribution $\in\Lambda( \lambda_P\geq 1/3, \lambda_D\geq 2/3-\lambda_P)$ is characterized as follows  

\begin{eqnarray}
\mathcal{D}(\Lambda( \lambda_P\geq 1/3, \lambda_D\geq 2/3-\lambda_P)\triangleq \bigg \{ \left( d_{11}, d_{12},d_{21},d_{22}\right)  \in \mathbb{R}_+^4:\\ \nonumber
&& d_{11}+d_{12}+d_{21}\leq 1\\ \nonumber 
&& d_{11}+d_{12}+d_{22}\leq 1\\ \nonumber
&& d_{11}+d_{21}+d_{22}\leq 1\\ \nonumber 
&& d_{12}+d_{21}+d_{22}\leq 1\bigg \}. 
\end{eqnarray}

\begin{proof}
The converse proof of the outer bounds is directly implied here and therefore is omitted. The achievability arguments are proved as follows. Let $\mathcal{D}_{'}$ be the degrees of freedom region for the 2-user SISO X channel with alternating CSIT under any distribution $\in\Lambda( \lambda_P\geq 1/3, \lambda_D\geq 2/3-\lambda_P)$. In order to fully characterize $\mathcal{D}_{'}$, we need to prove that $\mathcal{D}_{'}\equiv \mathcal{D}$. 
The points $K=(1,0,0,0)$, $L=(0,1,0,0)$, $M=(0,0,1,0)$, $N=(0,0,0,1)$ are the corner points of $\mathcal{D}$ and can be verified to belong to $\mathcal{D}_{'}$ through dedicating only one transmitter to send all the information symbols over all the time slots to only one receiver. While, $P=(1/3,1/3,1/3,1/3)$ lies in $\mathcal{D}^{'}$ through the achievability schemes of \textit{Theorem 1}. Now, consider any point $ d^{*}=(d_{11}, d_{12}, d_{21}, d_{22}) \in \mathcal{D}$ as defined by \textit{theorem 3}, we can easily verify that $d^{*}$ belongs to the convex hull whose corner points; K, L, M, N, P and $O=(0,0,0,0)$. In other words, we can show that $d^{*}$ is expressed as a convex combination of those points, K, L, M, N, P and O, of $\mathcal{D}^{'}$. For instance, $d^{*}=\alpha_1 K+\alpha_2 L+\alpha_3 M+\alpha_4 N+\alpha_5 O+ \alpha_6 P$, where the coefficients  $\alpha_i$ are non-negative for all $ d^{*} \in \mathcal{D}$ and their sum is one. The coefficients $\alpha_i$ defined in Table~\ref{Convex combination coefficients}, satisfy the two conditions of the convex combination and thereby all points  in $\mathcal{D}$ are convex combinations of the achievable points  K, L, M, N, P and O. Since the convex combinations are achievable by time sharing between the achievable schemes of the corner points,K, L, M, N, P and O, this implies that $\mathcal{D}^{'}\equiv \mathcal{D}$.
\end{proof}

\renewcommand{\arraystretch}{1.3}
\begin{table}[!ht]
\centering
\small
\begin{tabular}{|l|l|l|l|l|l|l|}
  \hline
 & $\alpha_1$ & $\alpha_2$ & $\alpha_3$ &  $\alpha_4$ &  $\alpha_5$ & $\alpha_6$ \\
  \hline
 $\sum_{i, j}d_{ij} \leq 1$ &  $d_{11}$&  $d_{12}$ & $d_{21}$ & $d_{22}$ & $0$ & \tiny{ $1-\sum_{i, j}d_{ij}$}\\
 \hline
$\sum_{i, j}d_{ij} < 1$ & $\frac{d_{11}-d_{12}-d_{21}-d_{22}+1}{3}$& $\frac{d_{12}-d_{11}-d_{21}-d_{22}+1}{3}$ & $\frac{d_{21}-d_{11}-d_{12}-d_{22}+1}{3}$&$\frac{d_{22}-d_{11}-d_{12}-d_{21}+1}{3}$ & \tiny{$\sum_{i, j}d_{ij}-1$}&$0$\\
\hline
\end{tabular}\\
\caption {Convex combination coefficients}\label{Convex combination coefficients}
\end{table}

\subsubsection{Total sum of Degrees of Freedom}

While the total sum of degrees of freedom ($DoF$) is defined as the maximum weighted sum of $d_{ij}$ over all achievable degrees of freedom region $\mathcal{D}$, the outer bound on the sum degrees of freedom $DoF_{out}$ is defined as the maximum weighted sum of $d_{ij}$ over $\mathcal{D}_{out}$. The following theorem presents a tight outer bound $DoF_{out}(\Lambda)$ for the sum degrees of freedom $DoF$.\\

 \textit{Theorem 4 :}  $DoF(\Lambda) \leq DoF_{out}(\Lambda) \triangleq \smash{\displaystyle\max_{\mathcal{D}_{out}}}(d_{11}, d_{12},d_{21}, d_{22}) = 4/3$, where $\Lambda( \lambda_P\geq 1/3, \lambda_D\geq 2/3-\lambda_P)$\\

\begin{proof}

 We formalize maximizing a weighted sum of $d_{ij}$ over $\mathcal{D}_{out}(\Lambda)$ as a linear programming problem. We
explicitly evaluate all the extreme points of the feasible space bounded by conditions in \textit{Theorem 2}; $\mathcal{D}_{out}(\Lambda)$,typically for linear programming problem the solution is one of the vertices of the feasible set i.e. the vertices of the $\mathcal{D}_{out}(\Lambda)$,
calculate the objective value -$DoF(\Lambda)$- at the extreme points(vertices), after eliminating the redundant bounds. The surprising finding is that the corresponding region of $d_{ij}$ to the optimal $DoF = DoF_{out}=4/3 $ is only one point $(1/3,1/3,1/3,1/3)$. The achievability of that point is illustrated in section \ref{TwoUserSchemes}.  
\end{proof}

\begin{figure}
  \centering
\includegraphics[width=.7\linewidth,height=.4\textheight]{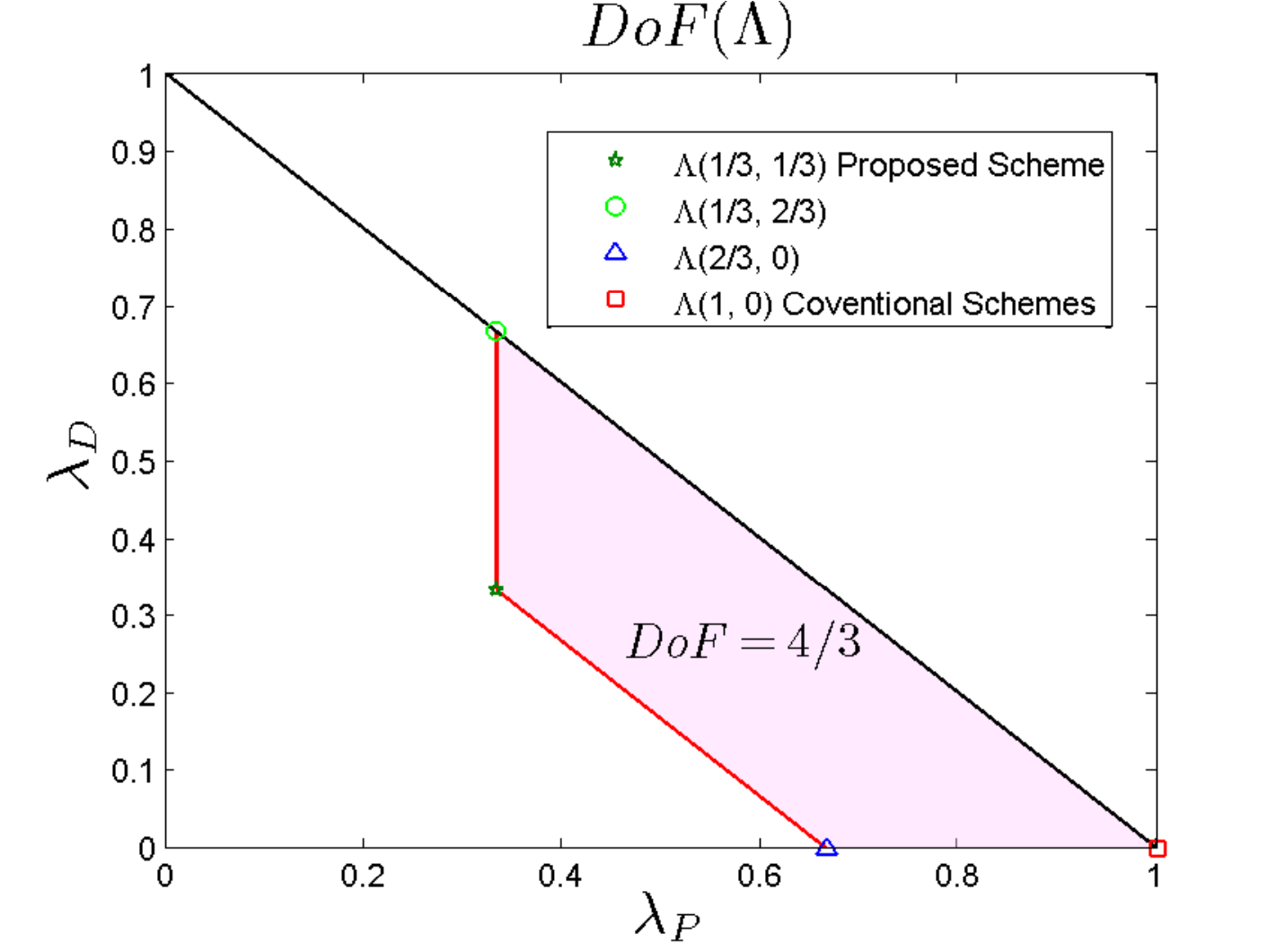} 
\caption{ Trade off between perfect and delayed CSIT \label{fig 4}}
\end{figure}

\textit{Remark 1:} \textbf{[Synergy benefits]}Note that the DoF for two-user X channel with perfect CSIT is
$4/3$ \cite{maddah2008communication, 4418479}, with delayed CSIT
is upper bounded by $6/5$ \cite{ghasemi2011degrees} and with No-CSIT is degraded to be unity due to the statistical independence of the two receivers in the total absence of channel state information at the transmitters \cite{6205390}. Synergy is the interaction
of multiple elements in a system to produce an effect greater than
the sum of their individual effects. In Particular, the
alternation of CSIT states $S_{ij}$ over three time slots works
cooperatively to provide a DoF greater than the DoF of the sum of
their individual DoF for the same network.

As an example, let us consider the CSIT alternation pattern given
by $(DD,DD,PP)$. If there is no interaction between the three time
slots, the DoF that can be obtained are given by $\frac{2}{3}
\frac{6}{5} + \frac{1}{3} \frac{4}{3}= \frac{56}{45}$ which is
lower than the upperbound on the DoF of the channel. However,
using achievable scheme 2, we can get $4/3$ DoF for this case as
this CSIT pattern is higher than $(DN,ND,PP)$ in Table 1. This illustrates the synergistic benefit that can be obtained from the alternation of CSIT over the three time slots.

Note that not all combinations of CSIT states could provide
synergistic benefits or work together in a cooperative way. For
example, when perfect CSIT comes before delayed and no CSIT , it
loses its synergetic DoF gain; where, the DoF degrades to the sum
of the individual DoF each case. on the other hand, when perfect
CSIT comes after delayed and no CSIT, the synergy of the alternation appears.\\

\textit{Remark 2:} \textbf{[Potential of delayed followed by
perfect CSIT]} The extraordinary synergistic gain of delayed CSIT
followed by perfect CSIT lies in the ability to upgrade the X
channel to a broadcast channel with delayed CSIT. In particular,
when the delayed CSIT comes first it provides the transmitters
with delayed channel knowledge which combats the distributed
nature of the X channel and can be exploited in addition to the
perfect CSIT to provide one message to each receiver.\\

\textit{Remark 3:} \textbf{[Combined No-CSIT]} We note that the
synergy of alternating CSIT is lost when the network has combined
No-CSIT in any time slot. Intuitively, the uncertainty of
distributed channel unawareness is better than blindness of
combined complete ignorance. Physically, the strategy of
interference creation phase is to create interference for the
receiver who can provide the transmitters with CSIT either perfect
or delayed to enable the transmitters to reconstruct the
interference in the interference resurrection phase. Hence,
combined no-CSIT is useless in the interference creation phase as
it provides nothing to transmitters and hence the transmitters
blindly create interference. While, in the interference resurrection  phase, the minimum  CSIT required to successfully reconstruct one interference term, formerly created from distributed transmitters, is $PN$ or $NP$ thereby combined no-CSIT again is useless in the interference resurrection phase as it provides nothing to transmitters and hence the transmitters can not reconstruct the interference terms. As Martin Luther king said before
``Darkness cannot drive out darkness; only light can do that.''\\

\textit{Remark 4:} \textbf{[Redundant Knowledge]}
Unlike what have been thought in the literature that the achievability of the upper bound on the degrees of freedom of X channel requires global and  continuous prefect CSIT $\equiv \Lambda(1,0,0)$, The first condition for achievability in theory 1 clarify that it requires alternating CSIT with distribution $\Lambda(1/3,1/3,1/3)$. Fig. 1 illustrates that there is a huge redundant CSIT greedily consumed or misused in the conventional schemes. Hence,  the region bounded by the red and black bold borders in Fiq.1 represent all distribution that achieves $DoF=4/3$, while the minimum CSIT is $\Lambda(1/3,1/3,1/3)$, $\Lambda(1, 0 ,0 )$ has the highest redundant knowledge. However, our proposed scheme consumes the CSIT wisely and no overhead information are sent or received \ref{schemes}, we have no evident to evaluate that either  $\Lambda(1/3,1/3,1/3)$ is the the distribution of the minimum CSIT knowledge to achieve the upper bound on the degrees of freedom of X channel or not. Therefore, the distribution of the minimum CSIT knowledge to achieve the upper bound on the degrees of freedom of X channel is still unknown.

\section{K-user SISO X-Network}
Motivated by our previous work in \cite{WagdyISIT, WagdyICC}, we extend the interference creation-resurrection scheme tailored originally for the two-user SISO X-channel to K-user X channel under the model for the CSIT availability; synergistic alternating CSIT. 

In this section, we propose a transmission scheme for the K-user SISO X-channel. Similarly to \cite{WagdyISIT}, the transmission scheme involves two phases, namely, interference creation and interference resurrection. The basic idea behind the proposed achievable scheme inspired by Maddah Ali in\cite{maddah2012completely}. Specifically, in MSIO BC with delayed CIST, transmitting date symbols to one receiver, inherently implies  receiving interference at the other receivers. Due to delayed CSI, after some delay, the transmitter has access to both past CSI and transmitted symbols and hence it perfectly knows the whole past received terms at each receiver. Imagine if each interference term at a receiver is a useful piece of information for specific other receivers about their desired symbols. Therefore, re-transmission of each of such interference terms not only aligns the past interference at one receiver, but also provides another receivers with a desired piece of information. on the other hand, in the context of X-channel, due to the distributed nature of the transmitter, the network loses the privileged of the joint processing of the transmitted signals at the collocated transmitters. However, in \cite{WagdyISIT}, the authors showed that the distributed nature of X-channel can be defeated by exploiting the synergistic benefits of the alternating CSIT; the delayed state and the perfect state. Utilizing this idea, here we show that the K-user SISO X-channel can achieve at least $\frac{2K}{K+1}$ DoF.

Before we proceed to the K-user case, as an illustrative example, we show that for the 3-user SISO X channel with alternating CSIT of $\Lambda(1/3, 1/3, 1/3)$ can achieve $3/2$ DoF. Let $u_1$, $u_2$ and $u_3$ be three independent data symbol intended to
$R_1$ transmitted from $T_1$, $T_2$ and $T_3$, respectively. Also, let
$v_1$, $v_2$ and $v_3$ be three independent data symbols intended to $R_2$ transmitted from $T_1$, $T_2$ and $T_3$, respectively. Similarly , $p_1$, $p_2$ and $p_3$ be three independent data symbols intended to $R_3$ transmitted from $T_1$, $T_2$ and $T_3$, respectively. In the next subsections, we show that we can reliably transmit the three symbols $(u_1 ,u_2, u_3)$ to receiver 1, $(v_1 ,v_2, v_3)$ to
receiver $2$  and, finally, $(p_1 ,p_2, p_3)$ to
receiver $3$ in $6$ time slots. 

Let us consider the alternating CSIT pattern given by $S_{123}^6 =
(NDD, DND, DDN, PPN, PDP, DPP)$. Here, the delayed CSIT is distributed over three time slots. Consequently, the interference creation phase consumes three time slots while the interference resurrection phase is executed over the other three time slot. The proposed scheme is performed in two separate phases as follows. 

\textit{Phase one:}, interference creation, each time slot of this phase is dedicated to each receiver where the transmitters
transmit three different linear combinations of the desired
messages, one term to each receiver. Since
$S_{123}(1)=NDD$, the first time slot is designed such that
interference is created for $R_2$ and $R_3$, hence, $T_1$ transmits $u_1$, $T_2$ transmits $u_2$ and $T_3$. The received signals at $R_1$, $R_2$ and $R_3$ are given respectively by
\begin{eqnarray}
Y_1(1)&\!\!\!=\!\!\!&h_{11}(1)u_1 + h_{12}(1)u_2 + h_{13}(1)u_3\equiv
L_1^1(u_1,u_2,u_3)\\
Y_2(1)&\!\!\!=\!\!\!&h_{21}(1)u_1 + h_{22}(1)u_2 + h_{23}(1)u_3 \equiv I_2^1(u_1,u_2,u_3)\\
Y_3(1)&\!\!\!=\!\!\!&h_{31}(1)u_1 + h_{32}(1)u_2 + h_{33}(1)u_3 \equiv I_3^1(u_1,u_2,u_3)
\end{eqnarray}
Therefore, $R_1$ receives the linear combination
$L_1^1(u_1,u_2,u_3)$ of its desired signals, while $R_2$ and $R_3$ receives only interference terms; $I_2^1(u_1,u_2,u_3)$ and $I_3^1(u_1,u_2,u_3)$. Similarly, in the next
two time slots, $T_1$ transmits $v_1$, $T_2$ transmits $v_2$ and $T_3$  transmits $v_3$ in the second time slots while $T_1$ transmits $p_1$, $T_2$ transmits $p_2$ and $T_3$  transmits $p_3$ in the third time slot. Then, the received signals at $R_1$, $R_2$ and $R_3$ are given respectively by
\begin{eqnarray}
Y_1(2)&\!\!\!=\!\!\!&h_{11}(2)v_1 + h_{12}(2)v_2+ h_{13}(2)v_3 \equiv I_1^1(v_1,v_2,v_3)\\
Y_2(2)&\!\!\!=\!\!\!&h_{21}(2)v_1 + h_{22}(2)v_2+ h_{23}(2)v_3\equiv
I_2^1(v_1,v_2,v_3)\\
Y_3(2)&\!\!\!=\!\!\!&h_{31}(2)v_1 + h_{32}(2)v_2+ h_{33}(3)v_3 \equiv I_3^1(v_1,v_2,v_3)
\end{eqnarray}
where  $R_2$ receives the first linear combination
$L_2^1(v_1,v_2,v_3)$ of its desired signals, while $R_1$ and $R_3$ receives the first interference terms $I_2^1(v_1,v_2,v_3)$ and $I_2^1(v_1,v_2,v_3)$.
while in the third time slot, 
\begin{eqnarray}
Y_1(3)&\!\!\!=\!\!\!&h_{11}(3)p_1 + h_{12}(3)p_2+ h_{13}(3)p_3 \equiv I_1^1(p_1,p_2,p_3)\\
Y_2(3)&\!\!\!=\!\!\!&h_{21}(3)p_1 + h_{22}(3)p_2+ h_{23}(3)p_3 \equiv
I_2^1(p_1,p_2,p_3))\\
Y_3(3)&\!\!\!=\!\!\!&h_{31}(3)p_1 + h_{32}(3)p_2+ h_{33}(3)p_3 \equiv L_3^1(p_1,p_2,p_3)
\end{eqnarray}
where  $R_3$ receives the first linear combination
$L_3^1(p_1,p_2,p_3)$ of its desired signals, while $R_1$ and $R_2$ receives the first interference terms $I_1^1(p_1,p_2,p_3)$ and $I_2^1(p_1,p_2,p_3)$.
By the end of time slot three each receiver receives one linear combination term from it's intended message and as a by product the other two receivers receives two interference terms. Now, we have six interference terms available to the three receivers. In the interference resurrection phase we will utilize these interference terms to provide the receivers with sufficient information to successfully decode their messages, specifically, each receiver needs another two linear combinations. Trivially, it requires six time slots to deliver six independent linear combinations. However, as we will show in the following that it will takes only three time slots by using the interference resurrection; exploiting interference as a common messages. \\

\textit{Phase two:} In the fourth time slot, interference resurrection phase begins, the transmitters utilizes the channel knowledge in $PPN$ to reconstruct $I_2^1(u_1,u_2,u_3)$ at $R_2$ while reconstructing $I_1^1(v_1,v_2,v_3)$ at $R_1$. As a result, $R_1$ and $R_2$  receive their second linear combination terms $L_1^2(u_1, u_2, u_3)$and $L_2^2(v_1, v_2,v_3)$ while $R_3$ receives pure interference. In particular, The transmitted signals are given by: 
\begin{eqnarray}
X_1(4)&\!\!\!=\!\!\!&h_{21}^{-1}(4)h_{21}(1)u_1+h_{11}^{-1}(4)h_{11}(2)v_1\\
X_2(4)&\!\!\!=\!\!\!&h_{22}^{-1}(4)h_{22}(1)u_2+h_{12}^{-1}(4)h_{12}(2)v_2\\
X_3(4)&\!\!\!=\!\!\!&h_{23}^{-1}(4)h_{23}(1)u_3+h_{13}^{-1}(4)h_{13}(2)v_3
\end{eqnarray}
Note that the transmitted signals, in interference resurrection phase, are beam-formed signals-not random linear combinations like in interference creation phase-dependent on both the current channel knowledge and the outdated channel knowledge formerly received at interference creation phase. For an instance, to construct $X_1(4)$, $T_1$ utilizes the instantaneous knowledge of $h_11(4)$ and the delayed knowledge of $h_11(2)$.  

Therefore, the received signals at $R_1$, $R_2$ and $R_2$ are
given respectively by
\begin{eqnarray}
Y_1(4)&\!\!\!\equiv\!\!\!&I_1^1(v_1,v_2,v_3)+ L_1^2(u_1, u_2, u_3)\\
Y_2(4)&\!\!\!\equiv\!\!\!&I_2^1(u_1,u_2,u_3)+ L_2^2(v_1, v_2,v_3)\\
Y_3(4)&\!\!\!\equiv\!\!\!&I_3^2(u_1,u_2,u_3)+ I_3^2(v_1,v_2,v_3)
\end{eqnarray}

In the fifth time slot, interference resurrection phase for user one and user three begins, the transmitters utilizes the channel knowledge in $PNP$ to reconstruct $I_3^1(u_1,u_2,u_3)$ at $R_3$ while reconstructing $I_1^1(p_1,p_2,p_3)$ at $R_1$. As a result, $R_1$ receives it's third linear combination term $L_1^3(u_1, u_2, u_3)$ and $R_3$ receives it's second interference term $L_3^2(p_1, p_2,p_3)$  while $R_2$ receives pure interference. In particular, The transmitted signals are given by: 
\begin{eqnarray}
X_1(5)&\!\!\!=\!\!\!&h_{31}^{-1}(5)h_{31}(1)u_1+h_{11}^{-1}(5)h_{11}(3)p_1\\
X_2(5)&\!\!\!=\!\!\!&h_{32}^{-1}(5)h_{32}(1)u_2+h_{12}^{-1}(5)h_{12}(3)p_2\\
X_3(5)&\!\!\!=\!\!\!&h_{33}^{-1}(5)h_{33}(1)u_3+h_{13}^{-1}(5)h_{13}(3)p_3
\end{eqnarray}
 As a result, the received signals at $R_1$, $R_2$ and $R_2$ are
given respectively by:
\begin{eqnarray}
Y_1(5)&\!\!\!\equiv\!\!\!&I_1^1(p_1,p_2,p_3) + L_1^3(u_1, u_2, u_3)\\
Y_2(5)&\!\!\!\equiv\!\!\!&I_2^2(u_1,u_2,u_3)+ I_2^2(p_1, p_2,p_3)\\
Y_3(5)&\!\!\!\equiv\!\!\!&I_3^1(u_1,u_2,u_3)+ L_3^2(p_1,p_2,p_3)
\end{eqnarray}

In the sixth time slot, interference resurrection phase for usder two and user three begins, the transmitters utilizes the channel knowledge in $NPP$ to reconstruct $I_2^1(p_1,p_2,p_3)$ at $R_2$ while reconstructing $I_3^1(v_1,v_2,v_3)$ at $R_3$. As a result, $R_2$ and $R_3$ receive their second linear combination terms $L_2^2(v_1, v_2,v_3)$ and $L_3^2(u_1, u_2, u_3)$ while $R_1$ receives pure interference. In particular, The transmitted signals are given by: 
\begin{eqnarray}
X_1(6)&\!\!\!=\!\!\!&h_{31}^{-1}(6)h_{31}(2)v_1+ h_{21}^{-1}(6)h_{21}(3)p_1\\
X_2(6)&\!\!\!=\!\!\!&h_{32}^{-1}(6)h_{32}(2)v_2+ h_{22}^{-1}(6)h_{22}(3)p_2\\
X_3(6)&\!\!\!=\!\!\!&h_{33}^{-1}(6)h_{33}(2)v_3+ h_{23}^{-1}(6)h_{23}(3)p_3
\end{eqnarray}
 As a result, the received signals at $R_1$, $R_2$ and $R_2$ are
given respectively by
\begin{eqnarray}
Y_1(6)&\!\!\!\equiv\!\!\!&L_1^2(u_1,u_2,u_3)+ I_1^1(v_1,v_2,v_3) \\
Y_2(6)&\!\!\!\equiv\!\!\!&I_2^1(u_1,u_2,u_3)+ L_2^2(v_1, v_2,v_3)\\
Y_3(6)&\!\!\!\equiv\!\!\!&I_3^2(u_1,u_2,u_3)+ I_3^2(v_1,v_2,v_3)
\end{eqnarray}\\

\textit{Theorem 4:}
The DoF of the $K$-user SISO X channel with synergistic alternating CSIT under any distribution $\in\Lambda( \lambda_P\geq 1/3, \lambda_D\geq 2/3-\lambda_P)$ is Lower bounded as follows:  

\begin{eqnarray}
DoF^{X}_{K\times K}(\lambda_P\geq 1/3, \lambda_D\geq 2/3-\lambda_P)\geq \dfrac{K^2}{K+{K\choose 2}}=\dfrac{2K}{K+1} 
\end{eqnarray}

\begin{proof}

The transmission scheme starts with transmission of information symbols in phase one, the interference creation phase, in a certain way that guarantees to create reconstructable interference terms while providing receivers with linear combinations of their intended data symbols. This phase consumes $K$ time slots to deliver $K$ different linear combination of the data symbols to $K$ different receivers while creating $K*(K-1)$ reconstructable interference terms. In contrast, phase two, the interference creation phase, This phase consumes ${K\choose 2}$ time slots to delivers $K*(K-1)$ new linear combinations of the data symbols to the indented receivers in oreder to successfully decode $K^2$ data symbols.

\textit{Phase One: \textquotedblleft Interference
Creation\textquotedblright}: This phase is associated with the
delayed CSIT and might have one to $K$ sub-phases where each
sub-phase consumes one time slot. The number of sub-phases depends
on whether the delayed CSIT of the channels to the two receivers
occurs simultaneously or not. In the first case where the delayed
CSIT occurs in the same time slot, i.e., $S_{1\cdots K}= D\cdots D$, phase one has only one sub-phase in which all data symbols are greedily transmitted, thus \textit{interference creation} happens.
Consequently, each receiver has one equation consisting of $K$
terms, the first term is a linear combination from the desired
symbols while others are the interference term. On the other
hand, when the delayed CSIT do not occur simultaneously, i.e.,
$S_{1\cdots K}\in\{ D\cdots N, N\cdots D, D\cdots P, P\cdots D \}$, phase one includes number of 
sub-phases greater than one. Each sub-phase is dedicated to transmit the data symbols of one receiver. Consequently, each receiver has $K$
different equations over $K$ time slots, one of them is a linear
combination of the desired symbols without interference and the
others are interference terms only.\\

\textit{Phase Two: \textquotedblleft Interference
Resurrection\textquotedblright}: This phase is associated with
perfect CSIT. Similar to phase one, this phase might have one or
two sub-phases depending on whether the perfect the CSIT occurs
simultaneously or not. In this phase, the transmitters reconstruct
the old interference by exploiting the delayed CSIT received in
phase one. When the two transmitters have perfect CSIT
simultaneously, phase two has only one sub-phase in which the two
transmitters reconstruct the old interference received in phase
one. Then, the transmitters transmit two independent messages
exploiting the combined perfect CSIT. On the other hand, when the
perfect CSIT is distributed over two time slots, phase two
consists of two sub-phases where each sub-phase is dedicated to
resurrect the interference for one receiver. Unlike combined
perfect CSIT, transmitters consume two time slots to totally
reconstruct the old interference and provide new linear
combination of the desired symbols to the receivers.
\end{proof}

Note that this bound is tight for $K=2$, for which the two user X-channel achieves the upper bound on the DoF of $4/3$. However, the lower bound state in Theorem 4 does not scale with $K$, it is strictly better than the best known lower bound for the network with only delayed CSIT $\Lambda(0,1,0)$, DoF $\geq \frac{4}{3}-\frac{3}{2(2k-1)}$ for all values of $K$ \cite{ghasemi2011degrees}.

\section{$K\times 2$-user SISO X-Network}
In this section, we characterize the degrees of freedom of the $K\times 2$-user SISO X-Network with synergistic alternating CSIT. To this extent, first, we provide the achievability schemes to $3\times 2$ and $4\times 2$ SISO X channel as illustrative examples. Then, we generalize our acheivability scheme to the $K$-user case.\\

\begin{figure}
  \centering
\includegraphics[scale= 1.2]{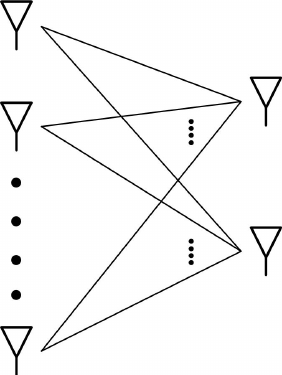} 
\caption{ $K\times2$-user SISO X network } 
\end{figure}

\textit{A.} Consider a $3\times 2$-user X network with synergistic alternation under certain distribution; $\Lambda(1/3, 1/3, 1/3)$.In order to achieve $4/3$ DoF, similar to the aforementioned schemes, the transmission strategy is executed in two distinctive phases; Interference creation and Interference resurrection phases, nevertheless, with minor modifications. Typically, the information symbols are avidity fed to some receivers in the interference creation phase in a form of random linear combinations while creating   interference to the other receivers. During the interference resurrection phase, the old interference terms, formerly created, are sent as a new linear combinations to the some receivers, ensuing interference alignment to other receivers. By the end, all the receivers have the desired number of equations in terms of their intended information symbols. In particular, to achieve $4/3$ DoF, We send six independent symbols to each receiver over nine time slots and  the alternating CSIT pattern is given by $S_{12}^9=(DD, DD, DD, PN,PN,PN, NP, NP, NP)$. Here, the delayed CSIT is collocated over three time slots. Consequently, the interference creation phase consumes three time slots to generate six linear combinations and six independent interference terms,three per each receiver, while the interference resurrection phase is executed over  the other six time slots to align the six interference terms formerly   created and ,as by product, generate new three linear combinations to each receiver. The proposed scheme is performed in two separate phases as follows. 

\textit{Phase one:}, interference creation, unlike previous interference creation phases in previous schemes, here, each time slot of this phase is dedicated to both the receivers ( send random linear combinations from the desired signals and create interference for both receiver simultaneously) where we send four different information symbols two for each receiver in each time slot. Since
$S_{12}^3=(DD, DD, DD)$, each time slot is designed such that
interference is created for $R_1$ and $R_2$, hence, $T_1$ transmits $u^1_1$ and $v^1_1$, $T_2$ transmits $v^1_2$, and $T_3$ transmits $u^1_3$. The received signals at $R_1$ and $R_2$ are given respectively by :
\begin{eqnarray}
Y_1(1)&\!\!\!=\!\!\!&h_{11}(1)u^1_1 + h_{13}(1)u^1_3 + h_{11}(1)v^1_1 + h_{12}(1)v^1_2 \\ &\!\!\!\equiv\!\!\!& L_1^1(u^1_1,u^1_3)+ I^1_1(v^1_1,v^1_2)\\
Y_2(1)&\!\!\!=\!\!\!&h_{21}(1)u^1_1 + h_{23}(1)u^1_3 + h_{21}(1)v^1_1 + h_{22}(1)v^1_2 \\ &\!\!\!\equiv\!\!\!& I^1_2(u^1_1,u^1_3)+ L^1_2(v^1_1,v^1_2)
\end{eqnarray}
Therefore, $R_1$ receives the first linear combination
$L^1_1(u^1_1,u^1_3)$ of its desired signals in addition to interference term $I^1_1(v^1_1,v^1_2)$, while $R_2$ receives the first linear combination $L^1_2(v^1_1,v^1_2)$ of its desired signals along with interference term; $I^1_2(u^1_1,u^1_3)$. Similarly, in the next two time slots, $T_1$ transmits $u^2_1$, $T_2$ transmits $u^1_2$ and $v^2_2$, and $T_3$ transmits $v^1_3$ in the second time slots while $T_1$ transmits $v^2_1$, $T_2$ transmits $u^2_2$ and $T_3$ transmits $u^2_3$ and $v^2_3$ in the third time slot. As a result, the received signals at $R_1$ and $R_2$ are given respectively by

\begin{eqnarray}
Y_1(2)&\!\!\!=\!\!\!&h_{11}(2)u^2_1 + h_{13}(2)u^1_2 + h_{11}(2)v^2_2 + h_{12}(2)v^1_3 \\ &\!\!\!\equiv\!\!\!& L_2^1(u^1_1,u^1_3)+ I^2_1(v^1_1,v^1_2)\\
Y_2(2)&\!\!\!=\!\!\!&h_{21}(2)u^2_1 + h_{23}(2)u^1_2 + h_{21}(2)v^2_2 + h_{22}(2)v^1_3 \\ &\!\!\!\equiv\!\!\!& I^2_2(u^1_2,u^2_1)+ L^2_2(v^1_3,v^2_2)\\
Y_1(3)&\!\!\!=\!\!\!&h_{12}(3)u^2_2 + h_{13}(3)u^2_3 + h_{11}(3)v^2_1 + h_{13}(3)v^2_3 \\ &\!\!\!\equiv\!\!\!& L^3_1(u^2_2,u^2_3)+ I^3_1(v^2_1,v^2_3)\\
Y_2(3)&\!\!\!=\!\!\!&h_{22}(3)u^2_2 + h_{23}(3)u^2_3 + h_{21}(3)v^2_1 + h_{23}(3)v^2_3 \\ &\!\!\!\equiv\!\!\!& I^3_2(u^2_2,u^2_3)+ L^3_2(v^2_1,v^2_3)
\end{eqnarray}
where  $R_3$ receives the first linear combination
$L_3^1(p_1,p_2,p_3)$ of its desired signals, while $R_1$ and $R_2$ receives the first interference terms $I_1^1(p_1,p_2,p_3)$ and $I_2^1(p_1,p_2,p_3)$.

By the end of time slot three each receiver receives three linear combination terms from it's intended information symbols along with  three interference terms. Now, we have six interference terms available to the two receivers. Then, the interference resurrection phase takes these interference terms to generate six common messages between the two receivers. Resurrecting interference terms is beneficial to the two receivers; one receiver utilize it by eliminating the interference terms from it's received signals in phase one while the other receiver receive it as a new linear combination from its information symbols. After, the interference resurrection phase the receivers has access to sufficient information to successfully decode their messages, specifically, each receiver needs six independent linear combinations from its information symbols.\\

\textit{Phase two:} In the fourth time slot, interference resurrection phase begins, the transmitters utilizes the channel knowledge in $PN$ to reconstruct $I_1^1(v^1_1,v^1_2)$ at $R_1$. As a result, $R_2$ receives the fourth linear combination term $L^4_2(v^1_1,v^1_2)$ while $R_1$ extracts its first linear combination term $L^1_1(u^1_1,u^1_3)$ by subtracting $Y_1(4)$ from $Y_1(1)$. In particular, The transmitted signals are given by: 
\begin{eqnarray}
X_1(4)&\!\!\!=\!\!\!&h_{11}^{-1}(4)h_{11}(1)v^1_1\\
X_2(4)&\!\!\!=\!\!\!&h_{12}^{-1}(4)h_{12}(1)v^1_2\\
\end{eqnarray}
Note that the transmitted signals, in interference resurrection phase, are beam-formed signals-not random linear combinations like in interference creation phase, dependent on both the current channel knowledge and the outdated channel knowledge formerly received at interference creation phase. For an instance, to construct $X_1(4)$, $T_1$ utilizes the instantaneous knowledge of $h_11(4)$ and the delayed knowledge of $h_11(1)$.  
Therefore, the received signals at $R_1$, $R_2$ and $R_2$ are
given respectively by
\begin{eqnarray}
Y_1(4)&\!\!\!\equiv\!\!\!&I_1^1(v^1_1,v^1_2)\\
Y_2(4)&\!\!\!=\!\!\!&h_{21}(4)h_{11}^{-1}(4)h_{11}(1)v^1_1+h_{22}(4)h_{12}^{-1}(4)h_{12}(1)v^1_2 \nonumber\\
&\!\!\!\equiv \!\!\!&L_4^2(v^1_1,v^1_2)
\end{eqnarray}

In the fifth time slot, interference resurrection phase for user 2, the transmitters utilizes the channel knowledge in $PN$ to reconstruct the interference term $I_2^1(u^1_1,u^1_3)$ at $R_2$. As a result, $R_2$ receives the same interference term received at time slot one and thereby $R_2$ extracts its first linear combination term $L^1_2(v^1_1, v^1_2)$ by subtracting $Y_5(3)$ from $Y_2(1)$ while, as a by product, $R_1$ receives it's fourth linear combination term $L_1^4(u^1_1, u^1_3)$. In particular, The transmitted signals are given by: 
\begin{eqnarray}
X_1(5)&\!\!\!=\!\!\!&h_{21}^{-1}(5)h_{21}(1)u^1_1\\
X_3(5)&\!\!\!=\!\!\!&h_{23}^{-1}(5)h_{23}(1)u^1_3
\end{eqnarray}
 As a result, the received signals at $R_1$ and $R_2$ are
given respectively by:
\begin{eqnarray}
Y_1(5)&\!\!\!=\!\!\!&h_{11}(5)h_{21}^{-1}(5)h_{21}(1)u^1_1 + h_{13}(5)h_{23}^{-1}(5)h_{23}(1)u^1_3\nonumber\\
&\!\!\!\equiv\!\!\!&L_1^4(u^1_1, u^1_3)\\
Y_2(5)&\!\!\!\equiv\!\!\!&I_2^1(u^1_1,u^1_3)
\end{eqnarray}

In the six time slot, the transmitters utilizes the channel knowledge in $PN$ to reconstruct $I_1^2(v^2_2,v^1_3)$ at $R_1$. As a result, $R_2$ receives the fifth linear combination term $L^5_2(v^2_2,v^1_3)$ while $R_1$ extracts its second linear combination term $L^2_1(u^2_1,u^1_2)$ by subtracting $Y_1(6)$ from $Y_1(2)$. In particular, The transmitted signals are given by: 
\begin{eqnarray}
X_2(6)&\!\!\!=\!\!\!&h_{12}^{-1}(6)h_{12}(2)v^2_2\\
X_3(6)&\!\!\!=\!\!\!&h_{13}^{-1}(6)h_{13}(2)v^1_3\\
\end{eqnarray}
Therefore, the received signals at $R_1$ and $R_2$ are
given respectively by:
\begin{eqnarray}
Y_1(6)&\!\!\!\equiv\!\!\!&I_1^2(v^2_2,v^1_3)\\
Y_2(6)&\!\!\!=\!\!\!&h_{22}(6)h_{12}^{-1}(6)h_{12}(2)v^2_2+h_{23}(6)h_{13}^{-1}(6)h_{13}(2)v^1_3 \nonumber\\
&\!\!\!\equiv \!\!\!&L_5^2(v^2_2,v^1_3)
\end{eqnarray}

In the seventh time slot, interference resurrection phase for user two, the transmitters utilizes the channel knowledge in $NP$ to reconstruct the interference term $I_2^2(u^2_1,u^1_2)$ at $R_2$. As a result, $R_2$ receives the same interference term received at time slot two and thereby $R_2$ able to extract its second linear combination term $L^2_2(v^2_2, v^1_3)$ by subtracting $Y_2(7)$ from $Y_2(2)$ while, as a by product, $R_1$ receives it's fifth linear combination term $L_1^5(u^2_1, u^1_2)$. In particular, The transmitted signals are given by: 
\begin{eqnarray}
X_1(7)&\!\!\!=\!\!\!&h_{21}^{-1}(7)h_{21}(2)u^2_1\\
X_2(7)&\!\!\!=\!\!\!&h_{22}^{-1}(7)h_{22}(2)u^1_2
\end{eqnarray}
 As a result, the received signals at $R_1$ and $R_2$ are
given respectively by:
\begin{eqnarray}
Y_1(7)&\!\!\!=\!\!\!&h_{11}(7)h_{21}^{-1}(5)h_{21}(1)u^2_1 + h_{12}(7)h_{22}^{-1}(7)h_{22}(1)u^1_2\nonumber\\
&\!\!\!\equiv\!\!\!&L_1^4(u^2_1, u^1_2)\\
Y_2(7)&\!\!\!\equiv\!\!\!&I_2^2(u^2_1,u^1_2)
\end{eqnarray}

In the eighth time slot, the transmitters utilizes the channel knowledge in $PN$ to reconstruct $I_1^2(v^2_1,v^2_3)$ at $R_1$. As a result, $R_2$ receives the sixth linear combination term $L^6_2(v^2_1,v^2_3)$ while $R_1$ extracts its third linear combination term $L^3_1(u^2_2,u^2_3)$ by subtracting $Y_1(8)$ from $Y_1(3)$. In particular, The transmitted signals are given by: 
\begin{eqnarray}
X_1(8)&\!\!\!=\!\!\!&h_{11}^{-1}(8)h_{11}(3)v^2_1\\
X_3(8)&\!\!\!=\!\!\!&h_{13}^{-1}(8)h_{13}(3)v^2_3\\
\end{eqnarray}
Therefore, the received signals at $R_1$ and $R_2$ are
given respectively by:
\begin{eqnarray}
Y_1(8)&\!\!\!\equiv\!\!\!&I_3^2(v^2_1,v^2_3)\\
Y_2(8)&\!\!\!=\!\!\!&h_{21}(8)h_{11}^{-1}(8)h_{11}(3)v^2_1+h_{23}(8)h_{13}^{-1}(8)h_{13}(3)v^2_3 \nonumber\\
&\!\!\!\equiv \!\!\!&L_6^2(v^2_1,v^2_3)
\end{eqnarray}

In the ninth time slot, interference resurrection phase for user two, the transmitters utilizes the channel knowledge in $NP$ to reconstruct the interference term $I_2^2(u^2_2,u^2_3)$ at $R_2$. As a result, $R_2$ receives the same interference term received at time slot two and thereby $R_2$ able to extract its third linear combination term $L^3_2(u^2_2, u^2_3)$ by subtracting $Y_2(9)$ from $Y_2(3)$ while, as a by product, $R_1$ receives it's sixth linear combination term $L_1^6(u^2_2, u^2_3)$. In particular, The transmitted signals are given by: 
\begin{eqnarray}
X_2(9)&\!\!\!=\!\!\!&h_{22}^{-1}(9)h_{22}(3)u^2_2\\
X_3(9)&\!\!\!=\!\!\!&h_{23}^{-1}(9)h_{23}(3)u^2_3
\end{eqnarray}
 As a result, the received signals at $R_1$ and $R_2$ are
given respectively by:
\begin{eqnarray}
Y_1(9)&\!\!\!=\!\!\!&h_{12}(9)h_{22}^{-1}(9)h_{22}(3)u^2_2 + h_{13}(9)h_{23}^{-1}(9)h_{23}(3)u^2_3\nonumber\\
&\!\!\!\equiv\!\!\!&L_1^4(u^2_2, u^2_3)\\
Y_2(9)&\!\!\!\equiv\!\!\!&I_3^2(u^2_2,u^2_3)
\end{eqnarray}
By the end of the ninth slot, each receiver has access to sufficient information to successfully decode it's symbols, specifically, each receiver has six independent equations (linear combinations)in six variables and six interference terms aligned in three dimensions.\\

\textit{B. Achievability scheme to even number of transmitters:}
 Consider a $4\times 2$-user X network with synergistic alternation under certain distribution; $\Lambda(1/3, 1/3, 1/3)$.In order to achieve $4/3$ DoF, similar to the aforementioned schemes, the transmission strategy is executed in two distinctive phases; Interference creation and Interference resurrection phases, nevertheless, without duplicating the number of transmitted symbols.  In particular, to achieve $4/3$ DoF, We send six independent symbols to each receiver over nine time slots and the alternating CSIT pattern is given by $S_{12}^9=(DD, DD, DD, PN,PN,PN, NP, NP, NP)$. Here, the delayed CSIT is collocated over two time slots. Consequently, the interference creation phase consumes two time slots to generate four linear combinations and four independent interference terms,three per each receiver, while the interference resurrection phase is executed over the other four time slots to align the four interference terms formerly created and ,as a by product, generate new two linear combinations to each receiver. The proposed scheme is performed in two separate phases as follows. 

\textit{Phase one:}, interference creation, unlike previous interference creation phases in previous schemes, here, each time slot of this phase is dedicated to both the receivers (send random linear combinations from the desired signals and create interference for both receiver simultaneously) where we send four different information symbols two for each receiver in each time slot. Since,
$S_{12}^2=(DD, DD)$, each time slot is designed such that
interference is created for $R_1$ and $R_2$, hence, $T_1$ transmits $u_1$, $T_2$ transmits $u_2$ and $v_2$, and $T_3$ transmits $v_3$. The received signals at $R_1$ and $R_2$ are given respectively by :
\begin{eqnarray}
Y_1(1)&\!\!\!=\!\!\!&h_{11}(1)u_1 + h_{12}(1)u_2 + h_{12}(1)v_2 + h_{13}(1)v_3 \\ &\!\!\!\equiv\!\!\!& L_1^1(u_1,u_2)+ I^1_1(v_2,v_3)\\
Y_2(1)&\!\!\!=\!\!\!&h_{21}(1)u_1 + h_{22}(1)u_2 + h_{22}(1)v_2 + h_{23}(1)v_3 \\ &\!\!\!\equiv\!\!\!& I^1_2(u_1,u_2)+ L^1_2(v_2,v_3)
\end{eqnarray}
Therefore, $R_1$ receives the first linear combination
$L^1_1(u_1,u_2)$ of its desired signals in addition to interference term $I^1_1(v_2,v_3)$, while $R_2$ receives the first linear combination $L^1_2(v_2,v_3)$ of its desired signals along with interference term; $I^1_2(u_2,u_2)$. Similarly, in the next time slots, $T_3$ transmits $u_3$, $T_4$ transmits $u_4$ and $v_4$, and $T_1$ transmits $v_1$. As a result, the received signals at $R_1$ and $R_2$ are given respectively by

\begin{eqnarray}
Y_1(2)&\!\!\!=\!\!\!&h_{13}(2)u_3 + h_{14}(2)u_4 + h_{11}(2)v_1 + h_{14}(2)v_4 \\ &\!\!\!\equiv\!\!\!& L_2^1(u_3,u_4)+ I^2_1(v_1,v_4)\\
Y_2(2)&\!\!\!=\!\!\!&h_{23}(2)u_3 + h_{24}(2)u_4 + h_{21}(2)v_1 + h_{24}(2)v_4 \\ 
&\!\!\!\equiv\!\!\!&I^2_2(u_3,u_4)+ L^2_2(v_1,v_4)
\end{eqnarray}
By the end of time slot two, end of interference creation phase, each receiver receives two linear combination terms from it's intended information symbols along with two interference terms. Now, we have four interference terms available to the two receivers. Then, the interference resurrection phase takes these interference terms to generate four common messages between the two receivers. Resurrecting interference terms is beneficial to the two receivers; one receiver utilize it by eliminating the interference terms from it's received signals in phase one while the other receiver receive it as a new linear combination from its information symbols. After, the interference resurrection phase the receivers has access to sufficient information to successfully decode their messages, specifically, each receiver needs four independent linear combinations from its information symbols.\\

\textit{Phase two:} In the third time slot, interference resurrection phase begins and extends for  four time slots, here we execute the interference resurrection phase in two separate stages as follows:

Stage one interference resurrection for $R_1$: the transmitters utilizes the channel knowledge in $PN$, received in time slot three and four, to reconstruct $I_1^1(v_2,v_3)$ and $I_1^1(v_1,v_4)$ at $R_1$. As a result, $R_2$ receives the third and fourth linear combination terms $L^3_2(v_2,v_3)$ and $L^4_2(v_1,v_4)$, respectively , while $R_1$ extracts its first and second linear combination terms $L^1_1(u_1,u_2)$ and $L^2_1(u_3,u_4)$ by subtracting $Y_1(3)$ from $Y_1(1)$ and $Y_1(4)$ from $Y_1(2)$, respectively. In particular, The transmitted signals are given by: 
\begin{eqnarray}
X_2(3)&\!\!\!=\!\!\!&h_{12}^{-1}(3)h_{12}(1)v_2\\
X_3(3)&\!\!\!=\!\!\!&h_{13}^{-1}(3)h_{13}(1)v_3\\
X_1(4)&\!\!\!=\!\!\!&h_{11}^{-1}(4)h_{11}(1)v_1\\
X_4(4)&\!\!\!=\!\!\!&h_{14}^{-1}(4)h_{14}(1)v_4\\
\end{eqnarray}
As a result, the received signals at $R_1$ and $R_2$, over the third and fourth time slots, are given respectively by:
\begin{eqnarray}
Y_1(3)&\!\!\!\equiv\!\!\!&I_1^2(v_2,v_3)\\
Y_2(3)&\!\!\!=\!\!\!&h_{21}(3)h_{12}^{-1}(3)h_{12}(3)v_2 + h_{23}(3)h_{13}^{-1}(3)h_{13}(2)v_3 \nonumber\\
&\!\!\!\equiv\!\!\!&L_2^3(v_2,v_3)\\
Y_1(4)&\!\!\!\equiv\!\!\!&I_1^2(v_1,v_4)\\
Y_2(4)&\!\!\!=\!\!\!&h_{21}(4)h_{11}^{-1}(4)h_{11}(2)v_1+h_{24}(4)h_{14}^{-1}(4)h_{14}(2)v_4 \nonumber\\
&\!\!\!\equiv \!\!\!&L_2^4(v_1,v_4)
\end{eqnarray}

Stage two interference resurrection for $R_2$: In the fifth and sixth time slots, the transmitters utilizes the channel knowledge in $PN$ to reconstruct the interference terms $I_2^1(u_1,u_2)$ and $I_2^2(u_3,u_4)$ at $R_2$. As a result, $R_2$ receives the same interference terms received at interference creation phase and thereby $R_2$ extracts its first and second linear combination term $L^1_2(v_2, v_3)$ and $L^2_2(v_1, v_4)$ by subtracting $Y_2(5)$ from $Y_2(1)$  and $Y_2(6)$ from $Y_2(2)$ while, as a by product, $R_1$ receives it's third and fourth linear combination terms $L_1^3(u_1, u_2)$ and $L_1^4(u_3, u_4)$(new information). In particular, the transmitted signals are given by: 
\begin{eqnarray}
X_1(5)&\!\!\!=\!\!\!&h_{21}^{-1}(5)h_{21}(1)u_1\\
X_2(5)&\!\!\!=\!\!\!&h_{22}^{-1}(5)h_{22}(1)u_2\\
X_3(6)&\!\!\!=\!\!\!&h_{23}^{-1}(6)h_{23}(1)u_3\\
X_4(6)&\!\!\!=\!\!\!&h_{23}^{-1}(6)h_{23}(1)u_4
\end{eqnarray}
 As a result, the received signals at $R_1$ and $R_2$, over the fifth and sixth time slots, are given respectively by:
\begin{eqnarray}
Y_1(5)&\!\!\!=\!\!\!&h_{11}(5)h_{21}^{-1}(5)h_{21}(1)u_1 + h_{12}(5)h_{22}^{-1}(5)h_{22}(1)u_2\nonumber\\
&\!\!\!\equiv\!\!\!&L_1^3(u_1, u_2)\\
Y_2(5)&\!\!\!\equiv\!\!\!&I_2^1(u_1,u_2)\\
Y_1(6)&\!\!\!=\!\!\!&h_{13}(6)h_{23}^{-1}(6)h_{23}(2)u_3 + h_{14}(6)h_{24}^{-1}(5)h_{24}(2)u_4\nonumber\\
&\!\!\!\equiv\!\!\!&L_1^4(u_3, u_4)\\
Y_2(6)&\!\!\!\equiv\!\!\!&I_2^2(u_3,u_4)
\end{eqnarray}

By the end of the sixth time slot, each receiver has access to sufficient information to successfully decode it's information symbols, specifically, each receiver has four independent equations (linear combinations)in four variables and four interference terms seized in only two dimensions.  

\textit{C. generalization to $K\times 2-user$ SISO X-Network}
in this subsection, we describe the extension to the interference creation-resurrection transmission strategy for the $K\times 2$-user SISO X-Channel with synergistic alternating CIST under $\Lambda(1/3,1/3, 1/3)$. The transmission scheme is a two-phase scheme, like the previous one, but with many stages in each phase. 
New random linear combination are sent to the receivers in phase one, interference creation phase, in a certain way that guarantee fed the receivers with certain number of equations of the information symbols as well as creating common messages between the receivers; in our case is the interference it self. phase two, interference resurrection, are responsible for delivering the common messages to the receivers and thereby providing each receiver with the required number of equations to successfully decode its intended information symbols. In the $K\times 2-user$ SISO X-Channel depicted at Fig.\ref{} each transmitter in the network has an independent message to be communicated to each receiver therefore the network has multiple of $2K$ independent messages communicating between its nodes. This directly implies that each receiver interest in decoding $K$ independent messages over the successful communication time (certain number of time slots of channel uses) consequently each receiver requires $K$ independent equations to resolve it's own messages.\\

\textit{Phase one:}In the interference creation phase, the $K$ transmitters send their messages in a certain way to provide each receiver with $K/2$ random linear combinations of $K$ information symbols corrupted by $K/2$ interference terms in $K/2$ time slots. Specifically, we divide the $2K$ information symbols available at transmitters into $K/2$ batches; each batch has four different symbols. In each batch, there are two different groups of two symbols, each group has symbols which are intended to certain receiver but generated at different transmitters. This strategy in diving the information symbols guarantee that the interference terms created in phase one are beneficial when resurrecting in phase two. Here, beneficial means that these interference terms can work as a common message for the two receivers. By the end  of phase one, each receiver has access to $K/2$ independent linear combinations of its own symbols corrupted with $K/2$ constituent(constructable and beneficial) interference terms.\\

\textit{Phase two:} In the interference resurrection phase, the transmitters utilize the delayed CSIT sent in phase one and the instantaneous CSIT to generate and broadcast common messages to the receiver by reconstructing the constituent interference terms formerly received in phase one. In particular, the transmitters generate and send $K$ common messages, $K/2$ messages for each receiver over two stages. Creating one common message(constituent interference term) directly implies providing one receiver with old interference term to extract new linear combination from an interference-corrupted linear combination formerly received in phase one while providing the other receiver with new linear combination.
Almost sure, all the transmitters has sufficient channel knowledge to create the common message(old constituent interference term) at certain receiver; the delayed CSIT received in phase one provide the transmitters with the old channel coefficient while the instantaneous CSIT enable them to nullify the effect of current channel coefficient and thereby the old interference term can resurrected. Sending such a common message only consume one time slot consequently the interference resurrection phase consumes $K$ time slots. By the end of phase two each receiver receives $K/2$ new linear combinations of its own symbols in a certain stage while receives $K/2$ constituent interference terms in the other stage, used to extract $K/2$ linear combinations. After delivering all these common messages, every receiver has access to $K$ linear combinations of its intended information symbols. It is proved in (----) that these $K$ linear combinations are linearly independent almost surely, and thus, each receiver can resolve all it's $K$ information symbols. Hence, the DoF of the $2 \times K$-user SISO X channel with synergistic alternating CSIT is lower bounded as follows:
\begin{eqnarray}
DoF^{X}_{2\times K}(\lambda_P\geq 1/3, \lambda_D\geq 2/3-\lambda_P)\geq \dfrac{2K}{K/2+K}=\dfrac{4}{3} 
\end{eqnarray}

We note that this lower bound is tight for K=2, for which the two-user X-channel achieve the upper bound on the DoF of $4/3$.

\section{$2\times K$-user SISO X-Network}
In this section, we characterize the degrees of freedom of the $2\times K$-user SISO X-Network with synergistic alternating CSIT. To this extent, first, we provide the achievability schemes to $2\times 3$ and  $2\times 4$  SISO X network as illustrative examples. Then, we generalize our acheivability scheme to the $2 \times K$-user case.\\

\begin{figure}
  \centering
\includegraphics[scale= 1.2]{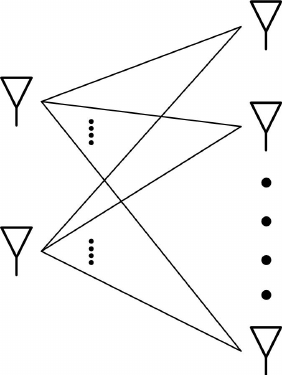} 
\caption{ $2\times K$-user SISO X network } \ref{2k-user}
\end{figure}

\textit{A.} Consider a $2\times 3$-user X network with synergistic alternation under certain distribution; $\Lambda(2/9, 2/9, 5/9)$. In order to achieve $4/3$ DoF, similar to the aforementioned schemes, the transmission strategy is executed in two distinctive phases; interference creation and interference resurrection phases, nevertheless, with minor modifications. In particular, to achieve $4/3$ DoF, specially in $2\times 3$-case, we send eight independent symbols; four symbols from each transmitter over six time slots and the alternating CSIT pattern is given by $S_{12}^6=(NDN, DNN, NND, NDN, PPN, NPP)$. Here, the delayed CSIT is distributive over four time slots. Consequently, the interference creation phase consumes four time slots to generate four linear combinations and four independent interference terms; one for $R_{1}$, two for $R_{2}$ and one for $R_{3}$,  while the interference resurrection phase is executed over the last two time slots to align the four interference terms formerly created and, as by product, generate new linear combinations to each receiver. The proposed scheme is performed in two separate phases as follows. 

\textit{Phase one:}, interference creation, unlike interference creation phases in $K\times 2$-user case, here, each time slot of this phase is dedicated to only one receiver where we send two different information symbols intended to certain receiver in each time slot. Since
$S_{12}^4=(NDN, DNN, NND, NDN)$, each time slot is designed such that interference is created for $R_1$ or $R_2$ or $R_3$ depending on the transmitted signal intended to which receiver, hence, in each time slot of the interference creation phase, $T_1$ and $T_2$ transmit information symbols intended to certain receiver. As a result, one receiver receive a linear combination from it's desired symbols (without interference) while the others receive only interference term. Form example, in time slot one, $T_1$ transmits $u^1_1$ and $T_2$ transmits $u^1_1$, consequently, the received signals are given respectively by :
\begin{eqnarray}
Y_1(1)&\!\!\!=\!\!\!&h_{11}(1)u^1_1 + h_{12}(1)u^1_2 \equiv L_1^1(u^1_1,u^1_2)\\
Y_2(1)&\!\!\!=\!\!\!&h_{21}(1)u^1_1 + h_{22}(1)u^1_2  \equiv I^1_2(u^1_1,u^1_2)\\
Y_3(1)&\!\!\!=\!\!\!&h_{31}(1)u^1_1 + h_{32}(1)u^1_2  \equiv I_3^1(u^1_1,u^1_2)
\end{eqnarray}

Therefore, $R_1$ receives the first linear combination
$L^1_1(u^1_1,u^1_2)$ of its desired signals, while $R_2$ and $R_3$ receive only interference terms $I^1_2(u^1_1,u^1_2)$ and $I_3^1(u^1_1,u^1_2)$, respectively. Similarly, in the next three time slots,   the transmitters send two independent linear combination to $R_{2}$ and one linear combination to $R_{3}$. In particular, $T_1$ transmits $v^1_1$ and $T_2$ transmits $v^1_2$ in the second time slots while, in third time slot, $T_1$ transmits $v^2_1$ and $T_2$ transmits $v^2_2$ after that, in the fourth time slot, the transmitter are dedicated to $R_3$ i.e. $T_1$ transmits $p^1_1$ and $T_2$ transmits $p^1_2$. As a result, the received signals at $R_1$, $R_2$ and $R_3$ are given respectively by:

\begin{eqnarray}
Y_1(2)&\!\!\!=\!\!\!&h_{11}(2)v^1_1 + h_{12}(2)v^1_2 \equiv I_1^1(v^1_1,v^1_2)\\
Y_2(2)&\!\!\!=\!\!\!&h_{21}(2)v^1_1 + h_{22}(2)v^1_2  \equiv L^1_2(v^1_1,v^1_2)\\
Y_3(2)&\!\!\!=\!\!\!&h_{31}(2)v^1_1 + h_{32}(2)v^1_2  \equiv I_3^1(v^1_1,v^1_2)\\
Y_1(3)&\!\!\!=\!\!\!&h_{11}(3)v^2_1 + h_{12}(3)v^2_2 \equiv L_1^2(v^2_1,v^2_2)\\
Y_2(3)&\!\!\!=\!\!\!&h_{21}(3)v^2_1 + h_{22}(3)v^2_2  \equiv L^2_2(v^2_1,v^2_2)\\
Y_3(3)&\!\!\!=\!\!\!&h_{31}(3)v^2_1 + h_{32}(3)v^2_2  \equiv I_3^2(v^2_1,v^2_2)\\
Y_1(4)&\!\!\!=\!\!\!&h_{11}(4)p^1_1 + h_{12}(4)p^1_2 \equiv I_1^1(p^1_1,p^1_2)\\
Y_2(4)&\!\!\!=\!\!\!&h_{21}(4)p^1_1 + h_{22}(4)p^1_2  \equiv I^1_2(p^1_1,p^1_2)\\
Y_3(4)&\!\!\!=\!\!\!&h_{31}(4)p^1_1 + h_{32}(4)p^1_2  \equiv L_3^1(p^1_1,p^1_2)
\end{eqnarray}

By the end of the fourth time slot receiver one and three receive two linear combination terms from their intended information symbols and four interference terms; two of four are useful interference terms and the others are useless ones. On the other hand, receiver two receives two linear combination terms from it's intended information symbols in addition to two interference terms all of them are useful terms. Now, we have two interference terms available to the second receiver in addition to two interference terms available to receiver one and three. Then, the interference resurrection phase takes the four interference terms to generate two common messages between the receivers. We note that resurrecting interference terms in each time slot is beneficial to two receivers only while the other one receive only interference. Contrary to the role of common messages in $K \times 2$-user scheme, one receiver utilize it by eliminating the interference terms from it's received signals in phase one while the other receiver receive it as a new linear combination from its information symbols, here, common messages are the sum two terms (old interference terms) where each receivers pair simultaneously receive it as new linear combination from their information symbols after eliminating the interference terms formerly received in the interference creation phase. After the interference resurrection phase, all receivers have access to sufficient information to successfully decode their messages.\\

\textit{Phase two:} In the fifth time slot, interference resurrection phase begins, the transmitters utilizes the channel knowledge of $PPN$, to simultaneously reconstruct $I_1^1(v^1_1,v^1_2)$ at $R_1$ and $I_2^1(u^1_1,u^1_2)$ at $R_2$. As a result, $R_2$ receives the third linear combination term $L^3_2(v^1_1,v^1_2)$ along with old interference term $I_2^1(u_1^1,u_2^1)$ while $R_1$ receives its second linear combination term $L^2_1(u^1_1,u^1_2)$ in addition to old interference term $I_1^1(v_1^1,v_2^1)$. In particular, The transmitted signals are given by: 
\begin{eqnarray}
X_1(5)&\!\!\!=\!\!\!&h_{21}^{-1}(5)h_{21}(1)u_1^1+h_{11}^{-1}(5)h_{11}(2)v_1^1\\
X_2(5)&\!\!\!=\!\!\!&h_{22}^{-1}(5)h_{22}(1)u_2^1+h_{12}^{-1}(5)h_{12}(2)v_2^2
\end{eqnarray}
Therefore, the received signals at $R_1$, $R_2$ and $R_2$ are given respectively by:
\begin{eqnarray}
Y_1(5)&\!\!\!=\!\!\!& L_1^2(u_1^1,u_2^1)+ I_1^1(v_1^1,v_2^1)\\
Y_2(5)&\!\!\!=\!\!\!& L_2^3(v_1^1,v_2^1)+ I_2^1(u_1^1,u_2^1)\\
Y_3(5)&\!\!\!=\!\!\!& I_3^3(v_1^1,v_2^1)+ I_3^2(u_1^1,u_2^1)\\
\end{eqnarray}

In the sixth time slot, we continue with interference resurrection phase for $R_2$ and $R_3$, similarly the transmitters utilizes the channel knowledge in $NPP$ to make interference resurrection simultaneously possible for $R_2$ and $R_3$,in particular, reconstructing the interference term $I_2^1(p^1_1,p^1_2)$ at $R_2$ and $I_3^1(v^2_1,v^2_2)$ at $R_3$. As a result, $R_2$ receives the fourth linear combination term $L^4_2(v^2_1,v^2_2)$ along with old interference term $I_2^1(p_1^1,p_2^1)$ while $R_3$ receives its second linear combination term $L^2_3(p^2_1,p^1_2)$ in addition to old interference term $I_3^2(v_1^2,v_2^2)$. In particular, The transmitted signals are given by:
\begin{eqnarray}
X_1(6)&\!\!\!=\!\!\!&h_{21}^{-1}(6)h_{21}(7)p_1^1+h_{31}^{-1}(6)h_{31}(3)v_1^2\\
X_2(6)&\!\!\!=\!\!\!&h_{22}^{-1}(6)h_{22}(7)p_2^1+h_{32}^{-1}(6)h_{32}(3)v_2^2
\end{eqnarray}
Therefore, the received signals at $R_1$, $R_2$ and $R_3$ are given respectively by:
\begin{eqnarray}
Y_1(6)&\!\!\!=\!\!\!& I_1^2(p_1^1,p_2^1)+ I_1^3(v_1^3,v_2^3)\\
Y_2(6)&\!\!\!=\!\!\!& L_2^4(v_1^2,v_2^2)+ I_2^1(p_1^1,p_2^1)\\
Y_3(6)&\!\!\!=\!\!\!& I_3^2(v_1^2,v_2^2)+ L_3^2(p_1^1,p_2^1)\\
\end{eqnarray}

By the end of the sixth slot, each receiver has access to sufficient information to successfully decode it's symbols, specifically, $R_1$ and $R_3$ have two independent equations (linear combinations)in two variables while $R_2$ has four independent equations in four variables and four interference terms aligned into two dimensions.\\

\textit{B. Achievability scheme to even number of receivers:}
 Consider a $2\times 4$-user X network with synergistic alternation under certain distribution; $\Lambda(1/6, 1/6, 4/6)$. In order to achieve $4/3$ DoF, similar to the previous scheme, the transmission strategy is executed in two distinctive phases; interference creation and interference resurrection phases, nevertheless, insert new variables $q_1$ and $q_2$ intended to $R_4$ instead of duplicating the number of transmitted symbols of the second receiver in the $2\times 3$-user scheme.  In particular, to achieve $4/3$ DoF, We send two independent symbols to each receiver over six time slots and the alternating CSIT pattern is given by $S_{12}^6=(NDNN, DNNN, NNDN, NNND, PPNN, NNPP)$. Here also, the delayed CSIT is distributed over four time slots. Consequently, the interference creation phase consumes four time slots to generate four linear combinations and twelve independent interference terms while the interference resurrection phase is executed over the last two time slots to align the four interference terms formerly created and, as a by product, generate new linear combinations to each receiver. The proposed scheme is exactly the same as the previous scheme in $2\times 3$-user case. We note that the required CSIT decreases with $K$, the number of receivers. 

\textit{C. generalization to $2\times K$-user SISO X-network}
in this subsection, we describe the extension to the interference creation-resurrection transmission strategy for the $2\times K$-user SISO X-Channel with synergistic alternating CIST under $\Lambda(2/3K, 2/3K, (3K-4)/3K)$. The transmission scheme is a two-phase scheme, like the previous one, but with many stages in each phase. 
New random linear combination are sent to the receivers in phase one, interference creation phase, in a certain way that guarantee fed the receivers with certain number of equations of the information symbols as well as creating common messages between the receivers; in our case is the interference it self. Phase two, interference resurrection, are responsible for delivering the common messages to the receivers and thereby providing each receiver with the required number of equations to successfully decode it's intended information symbols. In the $2\times K$-user SISO X-network depicted at Fig.\ref{2k_user} each transmitter in the network has an independent message to be communicated to each receiver therefore the network has multiple of $2K$ independent messages communicating between its nodes. This directly implies that each receiver interest in decoding $K$ independent messages over the successful communication time (certain number of time slots of channel uses) consequently each receiver requires $K$ independent equations to resolve it's own messages.\\

\textit{Phase one:}In the interference creation phase, the $K$ transmitters send their messages in a certain way to provide each receiver with $K/2$ random linear combinations of $K$ information symbols corrupted by $K/2$ interference terms in $K/2$ time slots. Specifically, we divide the $2K$ information symbols available at transmitters into $K/2$ batches; each batch has four different symbols. In each batch, there are two different groups of two symbols, each group has symbols which are intended to certain receiver but generated at different transmitters. This strategy in diving the information symbols guarantee that the interference terms created in phase one are beneficial when resurrecting in phase two. Here, beneficial means that these interference terms can work as a common message for the two receivers. By the end  of phase one, each receiver has access to only one linear combination of its own symbols as well as $(K-1)$ reconstructable interference terms.\\

\textit{Phase two:} In the interference resurrection phase, the transmitters utilize the delayed CSIT sent in phase one and the instantaneous CSIT to generate and broadcast common messages to the receivers by reconstructing the constituent interference terms formerly received in phase one. In particular, the transmitters generate and send $K$ common messages by adding $2K$ constituent interference terms generated in phase one. Creating one common message(constituent interference term) directly implies providing two receivers with old interference terms along with new linear combination.
Almost sure, all the transmitters has sufficient channel knowledge to create the common message(old constituent interference term) at certain receiver; the delayed CSIT received in phase one provide the transmitters with the old channel coefficient while the instantaneous CSIT enable them to nullify the effect of current channel coefficient and thereby the old interference term can resurrected. Sending such a common message only consume one time slot consequently the interference resurrection phase consumes $K/2$ time slots. By the end of phase $K$ receivers receive $K$ new linear combinations. After delivering all these common messages, every receiver has access to $2$ linear combinations of its intended information symbols. It is proved in (----) that these $2$ linear combinations are linearly independent almost surely, and thus, each receiver can resolve all it's $2$ information symbols. Hence, the DoF of the $2 \times K$-user SISO X channel with synergistic alternating CSIT is lower bounded as follows:
\begin{eqnarray}
DoF^{X}_{2\times K}(\lambda_P\geq 1/3, \lambda_D\geq 2/3-\lambda_P)\geq \dfrac{2K}{K/2+K}=\dfrac{4}{3} 
\end{eqnarray}

We note that this lower bound is tight for K=2, for which the two-user X-channel achieve the upper bound on the DoF of $4/3$.

\section{Prior art comparison and discussion}
\textit{A. Prior Art}
Here, in this subsection, we recall some prior art in different contexts which our transmission strategy (Interference creation resurrection) builds upon some of them and our achievability results have much in common when compared to their results. In the context of broadcast channel, Maddah-Ali et al. in \cite{Delayed_CSIT} have recently proposed an new multi-phase transmission scheme that can efficiently utilize the delayed CSIT in fast fading environment; channel coefficients are completely independent over time instances which means that knowing the past channel reveal nothing about the current channel. In particular, they showed that the $K$-user MISO broadcast channel with delayed CSIT can achieve nontrivial DoF of $\frac{K}{1+1/2+\cdots +1}\geq 1$. In addition, they proved that this is the upper bound to the DoF for that network. This significant gain in DoF is attributed to the intelligent transmission scheme of broadcasting a single of common interest to multiple receivers. The transmission scheme consists of $K$ constitutive phase where phase $k, 1\leq K \leq K$, broadcast order-$k$ common symbol and generate order $K+1$ common symbol to feed them to the next phase. The transmitters has the ability to effectively generate these common messages because of two main privileges, first, joint signals processing at the transmitters (broadcast channel with K antennas)and secondly the delayed CSIT. Exploiting both privileges, all the transmitters have the access to all information symbols intended to all users as well as all the channel coefficient of all the former time instances therefore the transmitters can easy construct the common symbols(what are called in \cite{Delayed_CSIT} overheard equations) moreover broadcast different linear combinations of them for different receivers in the same time slots.

On the other hand, in the context of networks of distributed transmitters like X channel, the delayed CSIT is not beneficial as it is in the broadcast channels (collocated transmitters) due to the loss of joint processing of signals at transmitters. In \cite{ghasemi2011degrees}, Ghasemi et. al. developed a new transmission strategy principally tailored for the SISO X channel to utilized the delayed CSIT in order to achieve DoF more than one. Interestingly, using that strategy, they showed that the delayed CSIT can provide DoF gain, particularly, the DoF of the $K$-user SISO X channel is $\frac{4}{3}-\frac{2}{3(3K-1)}$. However this result provide DoF gain bigger that the trivial one, it says that the DoF of the $K$-user SISO X-channel with delayed CSIT is lower bounded by a fixed number of $4/3$. There is a severe DoF loss due to the compound negative impact of the loss of joint signal processing at transmitters (Distributed transmitters) and the delayed CSIT compared to $\frac{K^2}{2K-1}$,the DoF of $K$-user SISO X channel with perfect CSIT, and $\frac{K}{1+1/2+\cdots+1/K}\approx \frac{K}{ln(K)}$, the DoF of $K$-user MSIO Broadcast channel with delayed CSIT. Nevertheless, this Ghasemi's scheme provide tight result of $6/5$ in the two-user case, $6/5$ proved to be the upper bound for the two-user X-channel with delayed CSIT in \cite{Linear_Delayed}, so far its results in the K-user case still unbounded above.

Using the same approach of creating common messages and broadcasting higher order-symbols, much research work in the literature proposed a new assumptions(works as catalyst in chemical reactions) in addition to the delayed CSIT assumption(main ingredient) in ordered to defeat the distributed nature of X channel and bridge that DoF gap. First, in \cite{Delayed_output}, the authors study the impact of what called Delayed Shannon feedback on the DoF of SISO X channel. The global delayed Shannon feedback is a feedback where all receivers feed the transmitters with the received signals (noisy channel outputs) in addition to the channel states (channel coefficient between each pair) after certain delay in other words each transmitter has access to all channel coefficient as well as all received signals. Under this strong assumption(huge knowledge accessible to transmitters), it is shown that the $K$-user SISO X channel with  global delayed Shannon feedback achieves $\frac{K}{1+\frac{1}{2}\cdots+\frac{1}{K}}$. On the other hand, under the partial delayed Shannon feedback(each transmitter receives delayed Shannon feedback for it own receiver pair in other words transmitter $k$ has access to the received signal of receiver $k$ as well as it's channel coefficient to receiver $k$), the sum degrees of freedom is lower bound by $\frac{2K}{K+1}$. 

Secondly, in\cite{GuideBT}, the authors proposed the use of relays to facilitate interference alignment and enhance the achievable DoF of distributed and fully-connected wireless networks, specifically, using a relay with multiple antennas in addition to the delayed CSIT thereby they managed to defeat the distributed nature of X-channel. In particular, they showed that the DoF of the $K$-user SISO X channel is $\frac{K}{1+\frac{1}{2} \cdots+\frac{1}{K}}$ with $K-$antenna half-duplex relay, where no CSIT at transmitters while global delayed CSI to the relay (means that the channels to the $K$-antenna relay and the transmitters are available to the relay after certain delay). However,in this model, the transmitters have no CSIT at all, it is reasonable to compare their results to the other results in the literature of X-channel with delayed CSIT as the relay has access to global delayed CSI.  \\

\textit{B. Numerical comparison}
In the lack of tight upper bounds for the $K-$user, $K\times 2$-user and $2\times K-$user SISO X-network under the alternating CSIT assumption, we can not claim any optimality to our achievablility schemes. However, in the following, we present a comprehensive comparison between our achievability schemes and the main schemes in literature of X networks. 
First, we begin with comparing our achievability schemes and results, under the alternating CSIT assumption, to their peers but under perfect CSIT. Indeed, we recall the results of \cite{X-Networks} which present the tightest upper bounds of $\frac{K^2}{2K-1}$ and $\frac{2K}{K+1}$ for the $K$-user and $2\times K$-user or $K \times 2$-user SISO X network with perfect CSIT, respectively. In addition to these upper bounds, the authors developed an asymptotic interference alignment scheme for the $M\times N$-user SISO X network ($K$-user when $M=N$), which partially align the interference to asymptotically achieve the previous upper bound within a constant gap, i.e. $\epsilon > 0$, over infinite channel extension. Specifically, they constructed an achievable scheme to achieve $(M-1)Nn^\Gamma + N(n+1)^\Gamma$ DoF over a $(M-1)Nn^\Gamma + N(n+1)^\Gamma$ symbol extension of the channel, where $\Gamma=(M-1)(N-1)$ so that the achievable DoF are arbitrary close to $\frac{MN}{M+N-1}$ when $n\longrightarrow \infty$. For sake of comparison, the number of time slot in the channel extension to achieve certain DoF, i.e., $\frac{2K}{K+1}$ DoF, the maximum achievable DoF for our scheme \ref{section_schems}, is a good metric of practical communication schemes however the partial interference alignment based scheme can more DoF. To achieve $\frac{2K}{K+1}$ DoF for the K-user SISO X network using the achievable scheme of \cite{X-Networks}, it requires sending $K(K-1)n^{(K-1)^2}+K(n+1)^{(K-1)^2}$ over a $(K-1)n^{(K-1)^2}+K(n+1)^{(K-1)^2}$ symbol channel extension and perfect CSIT while our scheme \ref{section_schems} simply send $K^2$ independent messages over $(K+{K \choose 2})$. As an example, for 3-user SISO X-channel, to achieve $3/2$ DoF, it requires almost sending $9365$ message over $6243$ time slots while our scheme requires sending $9$ messages over $6$ time slots. Moreover, our since the achievable
scheme essentially creates $K^2$ point-to-point links over a $(K+{K \choose 2})$ symbol extension of the channel, it provides an $\mathcal{O}(1)$ capacity characterization of the $K$-user X network while the asymptomatic interference alignment scheme
only yields a capacity characterization within $\mathcal{O}(log(SNR))$.

Secondly, in\cite{ghasemi2011degrees}, Ghasemi and et. al., showed the possibility of distributed retrospective interference alignment for the $K$-user SISO X network with delayed CSIT. They proposed a two-phase transmission scheme to tackle the main bottle nick of the distributed network; loss of joint signal processing of the signals at transmitters(each transmitter has access only to its own symbols). In particular, it proved that the K-user SISO X network can achieve $\frac{4}{3}-\frac{2}{3(3K-1)}$ DoF under the delayed CSIT assumption. However it was proved in \cite{Linear_Delayed} that this result is tight for $K=2$, there is no evidence so far to confirm that for$K>2$. Trivially, our achievablility result for the K-user case is strictly higher than this one, i.e., for  our case, the achievable DoF$\longrightarrow 2$ while for Ghassmi scheme tends to $4/3$ when $K\longrightarrow \infty$.\\

\text{C. Conjecture of DoF Scaling}
Our achievable degrees of freedom for the $K$-user, $K\times 2$-user and $2\times K$-user SISO X-network under the alternating CSIT assumption are tight for $K=2$, for which we achieve the upper bound of DoF of $4/3$ for the SISO X-channel with perfect CSIT. Moreover, in light of \textit{Remark 2}, the impact of synergy between instantaneous CSIT and delayed CSIT on enhancing the achievable degrees of freedom of the two user SISO X channel by defeating the distributed nature of the X channel in the transmitters side. In other words, this synergy enables virtual joint signal processing at transmitters, thereby upgrading the two-user X channel with synergistic alternating CSIT to two-user MISO broadcast channel with delayed CSIT. Therefore, these insights pose that the synergistic alternating CSIT, in terms of characterizing the DoF, could enhance the K-user SISO X network a K-user MISO broadcast channel with delayed CSIT as it is beneficial  in the two-user case. Consequently, we conjecture that, under the synergistic alternating CSIT with $\Lambda(1/3, 1/3, 1/3)$, the $K$-user SISO X-channel can achieve $\dfrac{K}{1+\frac{1}{2}}$, the upper bound on the degrees of freedom of the $K$-user MSIO broadcast channel with delayed CSIT. However, it does not seem to be possible through our achivability schemes (two-phase schemes), we believe that the multi-phase schemes have many things to offer to the $K$-user SISO X Networks with synergistic alternating CSIT. Indeed, multi-phase achievebility schemes are perfect match to the multi-interferer nature of $K$-user X networks.   

\section{Conclusion}
We investigated the synergistic benefits of alternating CSIT for the two-user X-channel and  K-user networks. Specifically, We proposed two-phase IA schemes and obtained new achievable results on the DoF of the two-user SISO X channel, $K\times 2$-user, $2\times K$-user and $K$-user SISO X-networks under synergistic alternating CSIT assumption. Most of our results are achievable DoF results. Even though an upper bound was shown to be tight in the two-user X-channel, it is not generally sufficient to characterize the outer bound on the DoF of the $K$-user case. It is worth mentioning that the
only known upper bound in the literature on the DoF of channels with delayed CSIT, to our knowledge, is
for the $K$-user MISO BC. However it is proved
to be tight, it's extension to other channels seems to be not a straightforward task, specially for X-networks due to the distributed nature of their  transmitter and receiver sides. An important future direction of this work is to develop new
upper bounds on the DoF of $K$-user X-networks studied here.

%

\bibliographystyle
{IEEEtran}
\vspace{5mm}
\bibliography{IEEEabrv,Nulls}

\end{document}